# GEOMETRIC QUANTIZATION BY PATHS

## PART I: THE SIMPLY CONNECTED CASE

PATRICK IGLESIAS-ZEMMOUR

ABSTRACT. For any connected and simply connected parasymplectic space $(\mathrm{X},\omega)$ with group of periods $\mathrm{P}_\omega \subsetneq \mathbf{R}$, we construct a prequantum groupoid $\mathbf{T}_\omega$ as a diffeological quotient of the space of paths in X. This object, built from the geometry of the classical system, serves as a unified structure for prequantization. The groupoid $\mathbf{T}_\omega$ has X as its objects, and its space of morphisms $\mathscr{Y}$ carries a canonical left-right invariant 1-form $\boldsymbol{\lambda}$ whose curvature encodes $\omega$. A key property is that the isotropy group at any point $x$, naturally arising as a quotient of the space of loops, is isomorphic to the torus of periods $\mathrm{T}_\omega = \mathbf{R}/\mathrm{P}_\omega$. Furthermore, the entire symmetry group $\mathrm{Diff}(\mathrm{X},\omega)$ acts as faithful automorphisms of $(\mathbf{T}_\omega,\boldsymbol{\lambda})$ without central extensions at this level. Built within the framework of diffeology, this construction generalizes classical prequantization by applying to broad classes of spaces, including those with singularities or infinite-dimensional aspects, and by accommodating generalized (e.g., irrational) tori of periods. This paper focuses on the simply connected case; the construction will be extended to general diffeological spaces in a subsequent publication.

## INTRODUCTION

Geometric quantization provides a powerful framework for quantizing classical mechanical systems described by symplectic manifolds. A fundamental step, known as prequantization, involves constructing a principal U(1)-bundle over the classical space of motions[1] X equipped with a connection whose curvature is the symplectic form $\omega$. This is classically possible if and only if the de Rham class $[\omega]$ satisfies an integrality condition.

---

*Date*: August 15, 2025.

2020 *Mathematics Subject Classification.* Primary 53D50, 58A05; Secondary 22A22, 55R65, 58A10.

Thanks to the Hebrew University of Jerusalem, Israel, for its continuous academic support. I am also grateful for the stimulating discussions and assistance provided by the AI assistant Gemini (Google).

[1]Following Souriau [Sou70], we use the term 'space of motions' for the base space describing the space of solutions of the classical dynamical system.

However, applying geometric quantization to more general spaces, particularly those arising from symplectic reduction which often possess singularities or are infinite-dimensional (such as loop spaces), presents significant challenges. Traditional bundle theory and differential geometry tools are heavily reliant on local Euclidean structure, which is absent in these generalized settings.

This paper presents a novel approach to prequantization by working within the framework of **diffeology**, a robust mathematical theory that provides tools for differential geometry on arbitrary sets, including spaces with singularities and infinite-dimensional spaces, without requiring local Euclidean charts.

Our core contribution is the construction of a **prequantum groupoid $\mathbf{T}_\omega$** directly from the geometry of the classical system $(\mathrm{X}, \omega)$. Unlike traditional approaches that build structures over the base space X (like principal bundles), our method is fundamentally **path-based**. For any connected and simply connected diffeological space X equipped with a closed 2-form $\omega$ having discrete periods,[2] we define $\mathbf{T}_\omega$ as a diffeological quotient of the space of paths in X.

This construction, rooted in the space of paths, inherently connects to Feynman's perspective on quantization, where paths are central.

The prequantum groupoid $\mathbf{T}_\omega$ serves as a **single, unified structure** for prequantization, replacing the classical principal bundle picture. It has X as its space of objects, and its space of morphisms $\mathscr{Y}$ carries a canonical left-right invariant 1-form $\boldsymbol{\lambda}$, the "prequantum potential," whose curvature encodes $\omega$.

Crucially, this path-based construction offers two key conceptual shifts and advantages:

* The quantum phase information, classically represented by the $\mathrm{U}(1)$ fiber, is not introduced externally but emerges naturally as the **isotropy group** at any point $x \in \mathrm{X}$. This isotropy group, arising as a quotient of the space of loops based at $x$, is isomorphic to the **torus of periods** $\mathrm{T}_\omega = \mathbf{R}/\mathrm{P}_\omega$. This reveals the quantum phase as an intrinsic property tied to the path geometry and the periods of $\omega$.
* The entire symmetry group of the classical system, $\mathrm{Diff}(\mathrm{X}, \omega)$ (the group of $\omega$-preserving diffeomorphisms), acts as **faithful automorphisms** of the *entire prequantum groupoid* $(\mathbf{T}_\omega, \boldsymbol{\lambda})$, without involving central extensions at this level. This provides a new perspective on the classical emergence of central extensions in geometric quantization, suggesting

[2] In diffeology, any strict subgroup $\mathrm{P}_\omega \subsetneq \mathbf{R}$ is a discrete diffeological space. This condition $\mathrm{P}_\omega \neq \mathbf{R}$ is necessary and sufficient for our construction and generalizes the classical integrality condition for manifolds.

that they arise when descending from this fundamental groupoid structure to structures (like bundles or section spaces) defined over the base space X.

By constructing the prequantum groupoid within the framework of diffeology, this approach generalizes classical prequantization to broad classes of spaces previously inaccessible to rigorous geometric methods, including those with singularities or infinite-dimensional aspects, and accommodates generalized (e.g., irrational) tori of periods. This also naturally recovers the classical prequantization scenario for manifolds where the period group $P_\omega$ is of the form $h\mathbf{Z}$ for Planck's constant $h$, which is the standard integrality condition.

This paper focuses on the connected and simply connected case, which allows for a clear development of the core construction and its fundamental properties. The construction will be extended to general diffeological spaces in a subsequent publication (Part II).

The paper is structured as follows: Section I reviews necessary concepts from diffeology. Section II states the main theorem, presenting the construction of the prequantum groupoid $\mathbf{T}_\omega$ and the prequantum 1-form $\boldsymbol{\lambda}$. Section III provides the detailed proof. Section IV discusses the symmetries of the prequantum structure. Finally, Section V offers remarks and applications, including interpretation, concrete examples (such as the exact case, the sphere, symplectic reduction, and loop spaces), and outlines directions for future research, particularly the connection to Feynman integrals and the generalization to non-simply connected spaces.

## I. Preliminaries

In this paper, we work within the framework of diffeological spaces, which provides a flexible setting for spaces potentially more general than smooth manifolds, finite or infinite dimensional, such as function spaces, quotient spaces, subspaces, spaces with singularities[3] etc.

We assume that the reader is familiar with the basics of diffeology: parametrizations, plots, diffeological spaces, smooth maps, inductions, subductions, diffeomorphisms, differential forms, etc. A comprehensive reference for diffeology is the book [PIZ13], or more precisely, the revised reprint recently published by the Beijing WPC [PIZ22], where misprints and some minor errors, as well as formatting flaws of the original version, have been corrected.

We will briefly recall some of the definitions that are not part of the basic framework and will be used in the following.

[3]The study of singular spaces by diffeology methods began with the article on the irrational torus $T_\alpha$, defined as the quotient of the 2-torus by an irrational flow of slope $\alpha \in \mathbf{R} - \mathbf{Q}$ [DI83].

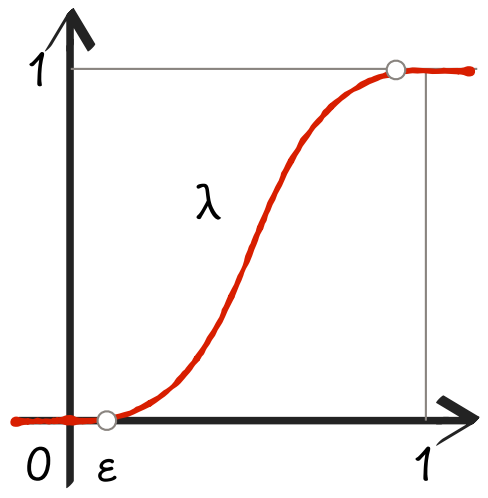

FIGURE 1. The smashing function λ.

**1. Paths and Stationary Paths.**

Let X be a diffeological space, we denote by

$$\mathrm{Paths}(\mathrm{X}) = \mathrm{C}^\infty(\mathbf{R}, \mathrm{X})$$

the space of smooth paths in X, equipped with the functionnal diffeology. For all $t \in \mathbf{R}$, we define $\hat t$ by:

$$\hat t \in \mathrm{C}^\infty(\mathrm{Paths}(\mathrm{X}), \mathrm{X}), \ \ \hat t(\gamma) = \gamma(t).$$

We denote by $\hat 0$ and $\hat 1$ the source and target maps. We denote by ends the *ends* map $\hat 0 \times \hat 1$ from Paths(X) to $\mathrm{X} \times \mathrm{X}$:

$$\mathrm{ends} \in \mathrm{C}^\infty(\mathrm{Paths}(\mathrm{X}), \mathrm{X} \times \mathrm{X}), \ \ \mathrm{ends} : \gamma \mapsto (\gamma(0), \gamma(1)).$$

We denote by $\mathrm{Paths}(\mathrm{X}, x, *)$, $\mathrm{Paths}(\mathrm{X}, *, x)$ and $\mathrm{Paths}(\mathrm{X}, x, x')$ the subspaces

$$\begin{aligned} \mathrm{Paths}(\mathrm{X}, x, *) &= \{\gamma \in \mathrm{Paths}(\mathrm{X}) \mid \hat 0(\gamma) = x\}, \\ \mathrm{Paths}(\mathrm{X}, *, x) &= \{\gamma \in \mathrm{Paths}(\mathrm{X}) \mid \hat 1(\gamma) = x\}, \\ \mathrm{Paths}(\mathrm{X}, x, x') &= \{\gamma \in \mathrm{Paths}(\mathrm{X}) \mid \mathrm{ends}(\gamma) = (x, x').\} \end{aligned}$$

We denote by Loops(X) the space of free loops of X,

$$\mathrm{Loops}(\mathrm{X}) = \{\ell \in \mathrm{Paths}(\mathrm{X}) \mid \hat 1(\ell) = \hat 0(\ell)\},$$

and we denote by $\mathrm{Loops}(\mathrm{X}, x)$ the set of Loops based at $x$,

$$\mathrm{Loops}(\mathrm{X}, x) = \mathrm{Paths}(\mathrm{X}, x, x) = \{\ell \in \mathrm{Loops}(\mathrm{X}) \mid \hat 0(\ell) = x\}.$$

We denote by comp the projection

$$\mathrm{comp} : \mathrm{Loops}(\mathrm{X}) \to \pi_0(\mathrm{Loops}(\mathrm{X})),$$

and also the projection from

$$\mathrm{comp} \restriction \mathrm{Loops}(\mathrm{X}, x) : \mathrm{Loops}(\mathrm{X}, x) \to \pi_0(\mathrm{Loops}(\mathrm{X}, x)) = \pi_1(\mathrm{X}, x).$$

Many constructions in homotopy involve concatenating paths, for $\gamma$ and $\gamma'$ when $\hat{1}(\gamma) = \hat{0}(\gamma')$, but the concatenation $\gamma \vee \gamma'$ is not necessarily smooth.

$$\gamma \vee \gamma' : t \mapsto \begin{cases} \gamma(2t) \text{ if } t \leq 1/2, \\ \gamma'(2t-1) \text{ if } t \geq 1/2, \end{cases}$$

This is why we consider the concatenation on the subspace of *stationary paths* [PIZ13, §5.4]. A path $\gamma \in \text{Paths}(X)$ is (strongly) stationary if there an $\varepsilon > 0$ such that:

$$\gamma \restriction ]-\infty, +\varepsilon[ = \gamma(0) \text{ and } \gamma \restriction ]1-\varepsilon, +\infty[ = \gamma(1).$$

Every path is fixed-ends homotopic to a stationary path by composing with the smashing function $\lambda$, $\gamma \mapsto \gamma^* = \gamma \circ \lambda$, see Figure 1. This is sufficient to ensure that many operations in diffeology involving the concatenation of paths, and depending only on their homotopy class, are justified.

**Convention.** To avoid unnecessary complications in the following, we shall consider only stationary paths without explicitly mentioning them. We will not change the notations so as not to make the text more cumbersome. That is, Paths(X) will denote the set of stationary paths, Loops(X) the space of stationary loops etc.

**2. Differential Forms.**

A differential $k$-form $\varepsilon$ on a diffeological space X, with $k \geq 0$, is a map $\varepsilon$ that associates to each $n$-plot $P : U \to X$ a smooth $k$-form $\varepsilon(P) \in C^\infty(U, \Lambda^k(\mathbf{R}^n))$, on $U = \text{dom}(P)$, where $\Lambda^k(\mathbf{R}^n)$ is the vector space of k-linear forms on $\mathbf{R}^n$, such that the following chain-rule is satisfied:

$$\varepsilon(P \circ F) = F^*\big(\varepsilon(P)\big),$$

for any $F \in C^\infty(V, U)$, where V is any Cartesian domain.[4] The space of differential $k$-forms on X is denoted by $\Omega^k(X)$. The space $\Omega^0(X)$ coincides with $C^\infty(X, \mathbf{R})$. Then, the exterior derivative is a linear operator defined by:

$$d : \Omega^k(X) \to \Omega^{k+1}(X), \text{ with } [d\varepsilon](P) = d[\varepsilon(P)].$$

The exterior derivative satisfies $d \circ d = 0$ and leads to the De Rham complex of closed and exact differential forms. The de Rham $k$-cohomology group is denoted by $\mathsf{H}^k_{\text{dR}}(X)$.

**Note on Terminology.** In this paper, a *parasymplectic form* on a diffeological space X is defined as a closed 2-form, $\omega \in \Omega^2(X)$ and $d\omega = 0$. The pair $(X, \omega)$ is called a *parasymplectic space.* No condition of non-degeneracy is assumed.

[4] A Cartesian domain is any open subset of some Cartesian space $\mathbf{R}^m$, for some $m$.

**3. The Chain-Homotopy Operator.**

Let X be a diffeological space and Paths(X) be equipped with the functional diffeology. Since X and Paths(X) are diffeological space, they have their own de Rham complex of differential forms denoted by $\Omega^k(\mathrm{X})$ or $\Omega^k(\mathrm{Paths}(\mathrm{X}))$, for the differential $k$-forms [PIZ13, §6.28]. There exists a smooth linear operator, called the *chain-homotopy operator*, defined in [PIZ13, §6.83]:

$$\mathsf{K} : \Omega^k(\mathrm{X}) \to \Omega^{k-1}(\mathrm{Paths}(\mathrm{X})),$$

satisfying the identity

$$\mathsf{K} \circ \mathrm{d} + \mathrm{d} \circ \mathsf{K} = \hat{1}^* - \hat{0}^*.$$

For $\alpha$ a $k$-form, it is explicitly given by this expression:

$$(\mathsf{K}\,\alpha)(\mathrm{P})_r(v_2)\cdots(v_k) = \int_0^1 \alpha\left(\begin{pmatrix} t \\ r \end{pmatrix} \mapsto \mathrm{P}(r)(t)\right)_{\binom{t}{r}} \begin{pmatrix} 1 \\ 0 \end{pmatrix}\begin{pmatrix} 0 \\ v_2 \end{pmatrix}\cdots\begin{pmatrix} 0 \\ v_k \end{pmatrix} dt,$$

where P is a plot in Paths(X), $r \in \mathrm{dom}(\mathrm{P})$ and $v_2 \cdots v_k$ are $(k-1)$ vectors at the point $r$.

**4. Groupoids.**

A diffeological groupoid $\mathbf{G}$ is defined by its space of objects $\mathrm{Obj}(\mathbf{G})$, which is a diffeological space, and a space of morphisms, or arrows, $\mathrm{Mor}(\mathbf{G})$, which is also a diffeological space, such that:

(1) The source and target maps $\mathrm{src}, \mathrm{trg} : \mathrm{Mor}(\mathbf{G}) \to \mathrm{Obj}(\mathbf{G})$ are smooth.
(2) The composition $(g, g') \mapsto g \cdot g'$, defined on the subset $\{(g, g') \in \mathrm{Mor}(\mathbf{G})^2 \mid \mathrm{trg}(g) = \mathrm{src}(g')\}$ of composable pairs, is smooth, where this subset is equipped with the subset diffeology of the product diffeology.
(3) The injection $x \mapsto \mathbf{1}_x$ from $\mathrm{Obj}(\mathbf{G})$ to $\mathrm{Mor}(\mathbf{G})$, that associates to each object $x \in \mathrm{Obj}(\mathbf{G})$ the identity $\mathbf{1}_x \in \mathrm{Mor}_{\mathbf{G}}(x, x)$, is an induction.

The latter condition ensures that the identity of $\mathbf{G}$ identifies diffeologically with the objects of $\mathbf{G}$. An important definition follows. Let $\mathrm{ends}(g) = (\mathrm{src}(g), \mathrm{trg}(g))$ denote the source/target map:

**Definition (Groupoid Automorphism).** *A* groupoid automorphism $\Phi$ *of* $\mathbf{G}$*, in diffeology, is a groupoid morphism from* $\mathbf{G}$ *to itself that is also a diffeomorphism on both the object and morphism spaces, generally denoted by* $(\Phi_{\mathrm{Mor}}, \Phi_{\mathrm{Obj}})$*, but most often in this paper as a pair* $(\Phi, \phi)$*. then:*

* $\Phi \in \mathrm{Diff}(\mathrm{Mor}(\mathbf{G}))$, $\phi \in \mathrm{Diff}(\mathrm{Obj}(\mathbf{G}))$.
* $\mathrm{ends} \circ \Phi = (\phi \times \phi) \circ \mathrm{ends}$, *with* $(\phi \times \phi)(x, x') = (\phi(x), \phi(x'))$.
* $\Phi(g \cdot g') = \Phi(g) \cdot \Phi(g')$.

*The group of automorphisms of* $\mathbf{G}$ *is denoted by* $\mathrm{Aut}(\mathbf{G})$.

The following definition [PIZ13, §8.4] is used to define the fiber bundles in diffeology.

**Definition (Fibrating Groupoid).** *A diffeological groupoid* $\mathbf{G}$ *is* fibrating *if the source/target map,* $\mathrm{ends} : \mathrm{Mor}(\mathbf{G}) \to \mathrm{Obj}(\mathbf{G}) \times \mathrm{Obj}(\mathbf{G})$*, is a subduction.*

### 5. Fiber Bundles and Principal fiber bundles.

Let $\pi : Y \to X$ be a map between two diffeological spaces. We build the characteristic groupoid $\pi$, with objects X and morphisms

$$\mathrm{Mor}(\pi) = \{\phi \in \mathrm{Diff}(Y_x, Y_{x'}) \mid (x, x') \in X \times X\},$$

with $Y_x = \pi^{-1}(x)$ equipped with the subset diffeology. This groupoid is equipped with a functional diffeology described in [PIZ85], and included in [PIZ13, §8.8].

**Definition (Fiber Bundle).** *The projection* $\pi$ *is a diffeological fibration if the characteristic groupoid* $\pi$ *is fibrating, that is, if the ends-map* $\phi \mapsto (x, x')$ *is a subduction.*

Principal fiber bundles are an important class of diffeological fiber bundles, because every fiber bundle in diffeology is associated with a principal bundle [PIZ13, §8.16]. We have the following theorem-definition [PIZ13, §8.11]:

**Definition (Principal Fiber Bundle).** *Let* G *be a diffeological group acting smoothly on a diffeological space* Y.[5] *Let* $\rho$ *be the graph of the action:*

$$\rho : Y \times G \to Y \times Y \;\; \textit{defined by} \;\; \rho : (y, g) \mapsto (y, g(y)).$$

*If* $\rho$ *is an induction then the projection* $\pi : Y \to X = Y/G$ *is a fiber bundle with fiber* G*, we call it a* principal fiber bundle *or a* principal fibration.

### 6. One-Dimensional Tori and Subgroups.

We call 1-dimensional torus any quotient $T = \mathbf{R}/\Gamma$, where $\Gamma \subsetneq \mathbf{R}$ is any strict subgroup of $\mathbf{R}$. Then,

**Proposition.** *Let* $H \subset T$ *be a subgroup, then, either* $H = T$ *or* H *is discrete.*

*Proof.* Let $\pi : \mathbf{R} \to T$ be the projection. Since $\pi$ is a homomorphism, the preimage $\pi^{-1}(H) \subset \mathbf{R}$ is a subgroup containing $\Gamma$. If $\pi^{-1}(H)$ contains an interval, then it coincides with $\mathbf{R}$ and $H = T$; this is a classic theorem. Otherwise, $\pi^{-1}(H)$ is a strict subgroup of $\mathbf{R}$, and according to [PIZ13, Ex. 124], it is discrete as a diffeological subgroup of $\mathbf{R}$. That means that its plots are locally constant. Now, let $P : U \to H$ be a plot. As a plot in T, it locally has a smooth lifting in $\mathbf{R}$ everywhere; let us say $Q : Y \to \mathbf{R}$ such that $\pi \circ Q = P \restriction V$. Thus, Q takes its values in $\pi^{-1}(H)$ which is discrete, then Q is locally constant, and so is P. Therefore, H is discrete. □

[5] Meaning the action map $\rho$ is smooth.

**7. Connections 1-form on Torus Principal Bundles.**

Let us consider a T-principal bundle $\pi : \mathrm{Y} \to \mathrm{X}$, where $\mathrm{T} = \mathbf{R}/\Gamma$ is a diffeological torus. For all $y \in \mathrm{Y}$ and $\tau \in \mathrm{T}$, let $\hat{y} : \mathrm{T} \to \mathrm{Y}$ be the orbit map $\hat{y}(\tau) = \tau(y)$.

**Definition (Connection 1-Form).** *A* connection form *on a* T*-principal bundle* $\pi : \mathrm{Y} \to \mathrm{X}$ *is any differential* 1*-form* $\lambda \in \Omega^1(\mathrm{Y})$, *such that:*

(1) *The form* $\lambda$ *is* invariant*: for all* $\tau \in \mathrm{T}$, $\tau_{\mathrm{Y}}^*(\lambda) = \lambda$.
(2) *The form* $\lambda$ *is* calibrated*: for all* $y \in \mathrm{Y}$, $\hat{y}^*(\lambda) = \theta$.

For details on general connections in diffeology see [PIZ13, §8.32], and specifically on T-principal bundles see [PIZ13, §8.37].

## II. Main Theorem: The Prequantum Groupoid

In this section, we present the construction of the prequantum groupoid associated with a closed 2-form on a simply connected diffeological space, as a pure diffeological quotient of the space of paths, without introducing external constructs. The assumption of simple connectedness simplifies key aspects, related to the triviality of fundamental groups, allowing for a focused development of the core groupoid structure and setting the stage for future generalization.

**8. The Prequantum Groupoid.**

Let X be a connected and simply connected diffeological space, and let $\omega$ be a closed 2-form on X:

$$\omega \in \Omega^2(\mathrm{X}) \ \text{ and } \ d\omega = 0.$$

For the remainder of this section, we assume $\omega \neq 0$.[6]

We use the chain-homotopy operator (see Article 3), $\mathsf{K} : \Omega^k(\mathrm{X}) \to \Omega^{k-1}(\mathrm{Paths}(\mathrm{X}))$, and consider $\mathsf{K}\omega \in \Omega^1(\mathrm{Paths}(\mathrm{X}))$.

Since $d\omega = 0$ and $\hat{1} \restriction \mathrm{Loops}(\mathrm{X}) = \hat{0} \restriction \mathrm{Loops}(\mathrm{X})$, the restriction of $\mathsf{K}\omega$ on Loops(X) is closed,

$$d[\mathsf{K}\omega \restriction \mathrm{Loops}(\mathrm{X})] = 0.$$

Thus,

**Definition (Group of Periods).** *For* $\omega$ *a closed* 2*-form on a simply connected space* X*, we define its* group of periods $\mathrm{P}_\omega$ *as the group of periods of the closed* 1*-form* $\mathsf{K}\omega \restriction \mathrm{Loops}(\mathrm{X})$.[7] *That is, it is the subgroup of* $(\mathbf{R}, +)$,

$$\mathrm{P}_\omega = \left\{ \int_\sigma \mathsf{K}\omega \mid \sigma \in \mathrm{Loops}(\mathrm{Loops}(\mathrm{X})) \right\},$$

[6] The case $\omega = 0$ is a degenerate situation resulting in the trivial pair groupoid $\mathbf{T}_0 = \mathrm{X}\times\mathrm{X}$ with the zero prequantum form $\boldsymbol{\lambda} = 0$, a case not relevant for the purpose of non-trivial prequantization.

[7] For a closed 2-form on a general diffeological space, the group of periods will be the group generated by the periods of all the restrictions of $\mathsf{K}\omega$ on the connected components of Loops(X).

Next, we consider the quotient group

$$\mathrm{T}_\omega = \mathbf{R}/\mathrm{P}_\omega \ \text{ and let } \ \pi_\omega : \mathbf{R} \to \mathrm{T}_\omega.$$

Following the definition of Article 6, $\mathrm{T}_\omega$ is a 1-dimensional diffeological group if $\mathrm{P}_\omega \neq \mathbf{R}$. It can be $\mathbf{R}$ if $\mathrm{P}_\omega = \{0\}$, or a circle $\mathrm{S}^1 \simeq \mathbf{R}/a\mathbf{Z}$ of perimeter $a$, if $\mathrm{P}_\omega$ has only one generator $a \in \mathbf{R}$, or an irrational torus otherwise. The fact that the quotient $\mathrm{T}_\omega$ is a 1-dimensional diffeological group is a property of the parasymplectic form $\omega$, or the space itself. This is why we introduce this definition:

**Definition (Torus of Periods).** *We shall say that the parasymplectic form* $\omega$ *is* discrete *if its group of periods* $\mathrm{P}_\omega$ *is a discrete subgroup of* $\mathbf{R}$*, or, which is equivalent, if* $\mathrm{P}_\omega \neq \mathbf{R}$*. In this case the quotient* $\mathrm{T}_\omega$ *is a* 1*-dimensional diffeological torus, we call it the* torus of periods.

We call it the torus of periods even if it is equal to $\mathbf{R}$ (a circle of inifinite perimeter). We say also that $\omega$ is *integral* when $\mathrm{P}_\omega = a\mathbf{Z}$, for some number $a$, then $\mathrm{T}_\omega \simeq \mathrm{S}^1$. In geometric quantization, Souriau defined a quantizable symplectic manifold $(\mathrm{M}, \omega)$ specifically when $\mathrm{P}_\omega = h\mathbf{Z}$, where $h$ is Planck's constant [Sou70].

The construction of the *prequantum groupoid* that follows applies only for discrete parasymplectic forms $\omega$.

**Remark.** Notice that this requirement for $\omega$ to be discrete ($\mathrm{P}_\omega \neq \mathbf{R}$) is the necessary and sufficient condition for the existence of the prequantum groupoid $\mathbf{T}_\omega$ as constructed here. This condition holds regardless of whether X is finite or infinite dimensional, or has singularities. It generalizes the classical integrality condition ($\mathrm{P}_\omega = a\mathbf{Z}$) typically required for the existence of a principal U(1)-bundle over a manifold. For second-countable Hausdorff manifolds, the period group $\mathrm{P}_\omega$ is always a discrete subgroup of $\mathbf{R}$ in the diffeological sense, meaning the diffeological discreteness condition $\mathrm{P}_\omega \neq \mathbf{R}$ is always fulfilled for such spaces.

Assume, then, that $\omega$ is discrete and X is simply connected.

The following relation defined on Paths(X) is the key to the construction of the prequantum bundle. Let $\gamma$ and $\gamma'$ be two paths in X, we say that

$$\gamma \sim_\omega \gamma' \ \text{ iff } \left\{ \begin{array}{l} \mathrm{ends}(\gamma) = \mathrm{ends}(\gamma') = (x, x'), \\ \exists [s \mapsto \gamma_s] \in \mathrm{Paths}(\mathrm{X}, x, x') : \gamma_0 = \gamma, \gamma_1 = \gamma', \\ \displaystyle\int_{[s \mapsto \gamma_s]} \mathsf{K}\omega = \int_0^1 \mathsf{K}\omega(s \mapsto \gamma_s)_s(1)\, ds \in \mathrm{P}_\omega. \end{array} \right.$$

We have, then:

**Proposition 1.** *The relation* $\sim_\omega$ *is an equivalence relation on* Paths(X).

We denote the quotient space by

$$\mathscr{Y} = \mathrm{Paths}(\mathrm{X})/\sim_\omega, \ \text{ and } \ \mathrm{class}_\omega : \mathrm{Paths}(\mathrm{X}) \to \mathscr{Y}.$$

The map ends : Paths(X) $\to$ X $\times$ X factors through $\mathcal{Y}$, defining the source and target maps for elements in $\mathcal{Y}$: $\mathrm{ends}([\gamma]_\omega) = \mathrm{ends}(\gamma)$.

$$\begin{array}{ccc} \mathrm{Paths}(X) & \xrightarrow{\ \mathrm{class}_\omega\ } & \mathcal{Y} \\ & {\scriptstyle \mathrm{ends}} \searrow \quad \swarrow {\scriptstyle \mathrm{ends}} & \\ & X \times X & \end{array}$$

We have then the main theorem of this construction:

**Theorem 2.** *Let* X *be a connected and simply connected diffeological space, and let* $\omega$ *be a closed* 2*-form on* X.

(1) *The space* $\mathcal{Y}$ *is the space of morphisms of a fibrating diffeological groupoid* $\mathbf{T}_\omega$*, which has* X *as its objects, i.e.,*

$$\mathrm{Obj}(\mathbf{T}_\omega) = X, \quad \textit{and} \quad \mathrm{Mor}(\mathbf{T}_\omega) = \mathcal{Y}.$$

*The source and target maps are given by* ends : $\mathcal{Y} \to X \times X$. *The groupoid composition is defined by the concatenation of paths:*

$$[\gamma]_\omega \cdot [\gamma']_\omega = [\gamma \vee \gamma']_\omega,$$

*for composable paths* $\gamma, \gamma'$*, and the inverse is*

$$[\gamma]_\omega^{-1} = [\bar\gamma]_\omega, \ \textit{with} \ \bar\gamma(t) = \gamma(1-t).$$

(2) *The isotropy group at any point* $x \in X$ *is isomorphic to the torus of periods:*

$$\mathbf{T}_{\omega,x} = \mathrm{Loops}(X, x)/\sim_\omega \simeq T_\omega.$$

(3) *There exists a unique differential* 1*-form*

$$\lambda \in \Omega^1(\mathcal{Y}) \ \textit{such that} \ \mathrm{class}_\omega^*(\lambda) = \mathsf{K}\omega.$$

*We call it the* prequantum 1-form *or* prequantum potential *of* $(X, \omega)$.

(4) *The* curvature *of* $\lambda$ *is related to* $\omega$ *by*

$$d\lambda + \mathrm{ends}^*(\omega \ominus \omega) = 0,$$

*where* $\omega \ominus \omega$ *is the* 2*-form* $\mathrm{src}^*(\omega) - \mathrm{trg}^*(\omega)$ *on* X $\times$ X, src *and* trg *being the source and target projections.*

(5) *The form* $\lambda$ *is invariant by left and right composition,*[8] *for any pair of composable arrows:*

$$\mathsf{L}(y)(y') = y \cdot y', \ \textit{and} \ \mathsf{R}(y)(y') = y' \cdot y.$$

*Meaning:*

$$\left\{ \begin{array}{lcl} \mathsf{L}(y)^*(\lambda \restriction \mathcal{Y}_{\mathrm{src}(y),*}) & = & \lambda \restriction \mathcal{Y}_{\mathrm{trg}(y),*}, \\ \mathsf{R}(y)^*(\lambda \restriction \mathcal{Y}_{*,\mathrm{trg}(y)}) & = & \lambda \restriction \mathcal{Y}_{*,\mathrm{src}(y)}, \end{array} \right.$$

[8] Also called precomposition and postcomposition.

**Definition (Prequantum Groupoid).** *The groupoid* $\mathbf{T}_\omega$ *is called the* prequantum groupoid *associated with* $\omega$. *The left-right invariant* 1*-form* $\boldsymbol{\lambda}$ *is called the* prequantum 1-form.

Also, since the prequantum groupoid is fibrating, see [PIZ13, §8.4], we have:

**Definition (Prequantum Bundles).** *For all* $x \in \mathrm{X}$, *let*

$$\hat{1}_x : \mathscr{Y}_x \to \mathrm{X}, \text{ defined by } \hat{1}_x : [\gamma]_\omega \mapsto \hat{1}(\gamma).$$

*These projections are all equivalent and are called* prequantum bundles *associated with* $\omega$. *The restriction of the prequantum* 1*-form* $\lambda = \boldsymbol{\lambda} \restriction \mathscr{Y}_x$ *becomes a connection* 1*-form for the action of* $\mathbf{T}_{\omega,x} \simeq \mathrm{T}_\omega$, *also called the* prequantum 1-form.

**Note 1.** When X is a manifold and $\mathrm{P}_\omega = a\mathbf{Z}$ for some $a > 0$, $\mathrm{T}_\omega \simeq \mathbf{R}/a\mathbf{Z} \simeq \mathrm{S}^1$, and the prequantum bundles built here are manifolds and are isomorphic to the principal $\mathrm{S}^1$-bundle of the classical prequantization construction [Sou70].

This construction, while restricted here to simply connected diffeological spaces, generalizes the classical construction in two directions: 1) It applies to any (simply connected) diffeological space, finite or infinite dimensional, with or without singularities. And 2) It extends the construction of a principal bundle to a groupoid, as the quotient of the space of paths, which includes, in particular, a natural source/target symmetry that has been broken in the prequantum bundle.

**Note 2.** A critical philosophical distinction of this construction lies in the emergent nature of the isotropy group $\mathbf{T}_{\omega,x} \simeq \mathrm{T}_\omega$, and this is probably the most remarkable aspect of this construct. Unlike classical geometric quantization, where the introduction of a U(1)-bundle (the "circle") often serves as a necessary mathematical device to overcome the "no-go theorem" (by ensuring the identity operator corresponds to multiplication by one), here the circle (or torus of periods) arises naturally as an intrinsic component of the path-space quotient itself, as the quotient of the space of loops. This suggests that the "quantum phase" is not a property added for technical consistency, but an inherent characteristic revealed through the reduction of paths. Precisely, the group of periods is a homomorphic representation of the $\pi_1$ of the space of loops.

**Note 3.** Let us anticipate a remark that is bound to come up when applying this construction to singular spaces. In the simply connected case, the space of loops is connected and its homotopy groups are conjugate. Consequently, there is no intrinsic reason within this framework for the quantum phase (isotropy), which is a quotient of the space of loops, to degenerate or vary across the dynamical parasymplectic simply connected space,[9] even at singular points. Its nature is fixed by the periods of $\omega$, consistent throughout the space. If quantum phenomena are related to such singularities, it would rather be through the representations of the group of automorphisms (which reveal the singular structure via the Klein stratification [PIZ13, §1.42]) than through the local isotropy itself.

[9]On a non-simply connected space, the homotopy of the space of loops is more complex but it seems that this is the aggregation of the groups of periods of $\mathsf{K}\omega$ on all the connected components

**Note 4.** Our construction of prequantum groupoid provides a fundamental object from which prequantum structures can be derived. It generalizes the setting where prequantum bundles exist for manifolds with any group of periods, as presented in [PIZ95]. In the general case of a non-simply connected diffeological space X, while the prequantum groupoid itself might be unique, this framework provides the necessary structure to investigate how the interplay between $\pi_1(X)$ and $P_\omega$ governs the types of prequantum bundles and representations that can be constructed over X from the groupoid, expected to be classified in a manner analogous to the prequantum bundles over manifolds by elements of $\mathrm{Ext}(\pi_1(X)^{\mathrm{ab}}, P_\omega)$ [PIZ95]. We will further explore the benefits of this approach.

## III. Proof of the Main Theorem

In this section, we will prove the theorem stated in Section II. The proof will be developed step-by-step, with each element established in successive articles. For technical convenience, the paths we consider are assumed to be stationary, allowing their concatenation to be a smooth path. This is a technical simplification without affecting the result, as the space of stationary paths is a deformation retract of the space of all smooth paths.

### 9. Integration of Closed 1-forms.

Let X be any connected diffeological space and $\alpha$ be a closed 1-form:

$$\alpha \in \Omega^1(X) \quad \text{and} \quad d\alpha = 0.$$

We define the *group of periods* of $\alpha$ by:

$$P_\alpha = \left\{ \int_\ell \alpha \mid \ell \in \mathrm{Loops}(X) \right\}.$$

The group of periods is a subgroup of $\mathbf{R}$. We say that $\alpha$ is *discrete* if $P_\alpha$ is a strict subgroup of $\mathbf{R}$, which is equivalent to $P_\alpha \neq \mathbf{R}$. Then, in this case, the *torus of periods*

$$T_\alpha = \mathbf{R}/P_\alpha$$

is a 1-dimensional diffeological group. The proposition [PIZ13, §8.28] asserts that:

$\mathrm{Loops}_i(\mathrm{Loops}(X))$ that will come into play for $P_\omega$. So, that will not change the situation: on a dynamical system the quantum phase is constant on connected components.

**Proposition.** *Let* X *be a connected diffeological space and* $\alpha$ *be a discrete closed* 1*-form. Then, there exists a smooth map* $f : \mathrm{X} \to \mathrm{T}_\alpha$*, unique up to a constant, such that*

$$\alpha = f^*(\theta),$$

*where* $\theta \in \Omega^1(\mathrm{T}_\alpha)$ *is the canonical* 1*-form, projection of* $dt \in \Omega^1(\mathbf{R})$ *by the projection* $\pi_\alpha : \mathbf{R} \to \mathrm{T}_\alpha$*. That is, defined uniquely by* $\pi_\alpha^*(\theta) = dt$*.*

We call this function the *integration function* of $\alpha$.

In the following, this construction will be applied to the subspace Loops(X) $\subset$ Paths(X) and the restriction of the 1-form $\mathsf{K}\omega \upharpoonright \mathrm{Loops(X)}$.

**10. Integrating concatenations over homotopies.**

Let X be a connected diffeological space, and let $\omega$ be a closed 2-form on X. All paths in the following are assumed to be stationary. Let $\gamma_0$ and $\gamma'_0$ be two paths such that $\gamma_0(1) = \gamma'_0(0)$. Then let

$$\sigma : s \mapsto \gamma_s,\ \sigma' : s \mapsto \gamma'_s, \text{ and } \sigma * \sigma' : s \mapsto \gamma_s \vee \gamma'_s,$$

where $\sigma$ is a homotopy from $\gamma_0$ to $\gamma_1$ and $\sigma'$ a homotopy from $\gamma'_0$ to $\gamma'_1$, such that $\gamma_s(1) = \gamma'_s(0)$ for all $s$. The homotopy $\sigma * \sigma'$ is the resulting homotopy from $\gamma_0 \vee \gamma'_0$ to $\gamma_1 \vee \gamma'_1$. Now, let $\mathsf{K}$ be the chain-homotopy operator, then:

$$\mathsf{K}\omega(\sigma * \sigma') = \mathsf{K}\omega(\sigma) + \mathsf{K}\omega(\sigma'), \text{ and } \int_{\sigma*\sigma'} \mathsf{K}\omega = \int_\sigma \mathsf{K}\omega + \int_{\sigma'} \mathsf{K}\omega.$$

*Proof.* By definition, $\int_{\sigma*\sigma'} \mathsf{K}\omega = \int_0^1 \mathsf{K}\omega(\sigma*\sigma')_t(1)\, dt$. Let us show the additivity

$$\mathsf{K}\omega(\sigma * \sigma')_t(1) = \mathsf{K}\omega(\sigma)_t(1) + \mathsf{K}\omega(\sigma')_t(1),$$

which will give, by integration, the identity we are looking for. From the definition of the chain-homotopy operator [PIZ13, §6.83], we have

$$\begin{aligned}
\mathsf{K}\omega(\sigma*\sigma')_t(1) &= \int_0^1 \omega\left(\begin{pmatrix} s \\ t \end{pmatrix} \mapsto (\sigma*\sigma')(t)(s)\right)_{\binom{s}{t}} \begin{pmatrix} 1 \\ 0 \end{pmatrix} \begin{pmatrix} 0 \\ 1 \end{pmatrix} ds \\
&= \int_0^1 \omega\left(\begin{pmatrix} s \\ t \end{pmatrix} \mapsto [\gamma_t \vee \gamma'_t](s)\right)_{\binom{s}{t}} \begin{pmatrix} 1 \\ 0 \end{pmatrix} \begin{pmatrix} 0 \\ 1 \end{pmatrix} ds \\
&= \int_0^{1/2} \omega\left(\begin{pmatrix} s \\ t \end{pmatrix} \mapsto \gamma_t(2s)\right)_{\binom{s}{t}} \begin{pmatrix} 1 \\ 0 \end{pmatrix} \begin{pmatrix} 0 \\ 1 \end{pmatrix} ds \\
&+ \int_{1/2}^1 \omega\left(\begin{pmatrix} s \\ t \end{pmatrix} \mapsto \gamma'_t(2s-1)\right)_{\binom{s}{t}} \begin{pmatrix} 1 \\ 0 \end{pmatrix} \begin{pmatrix} 0 \\ 1 \end{pmatrix} ds,
\end{aligned}$$

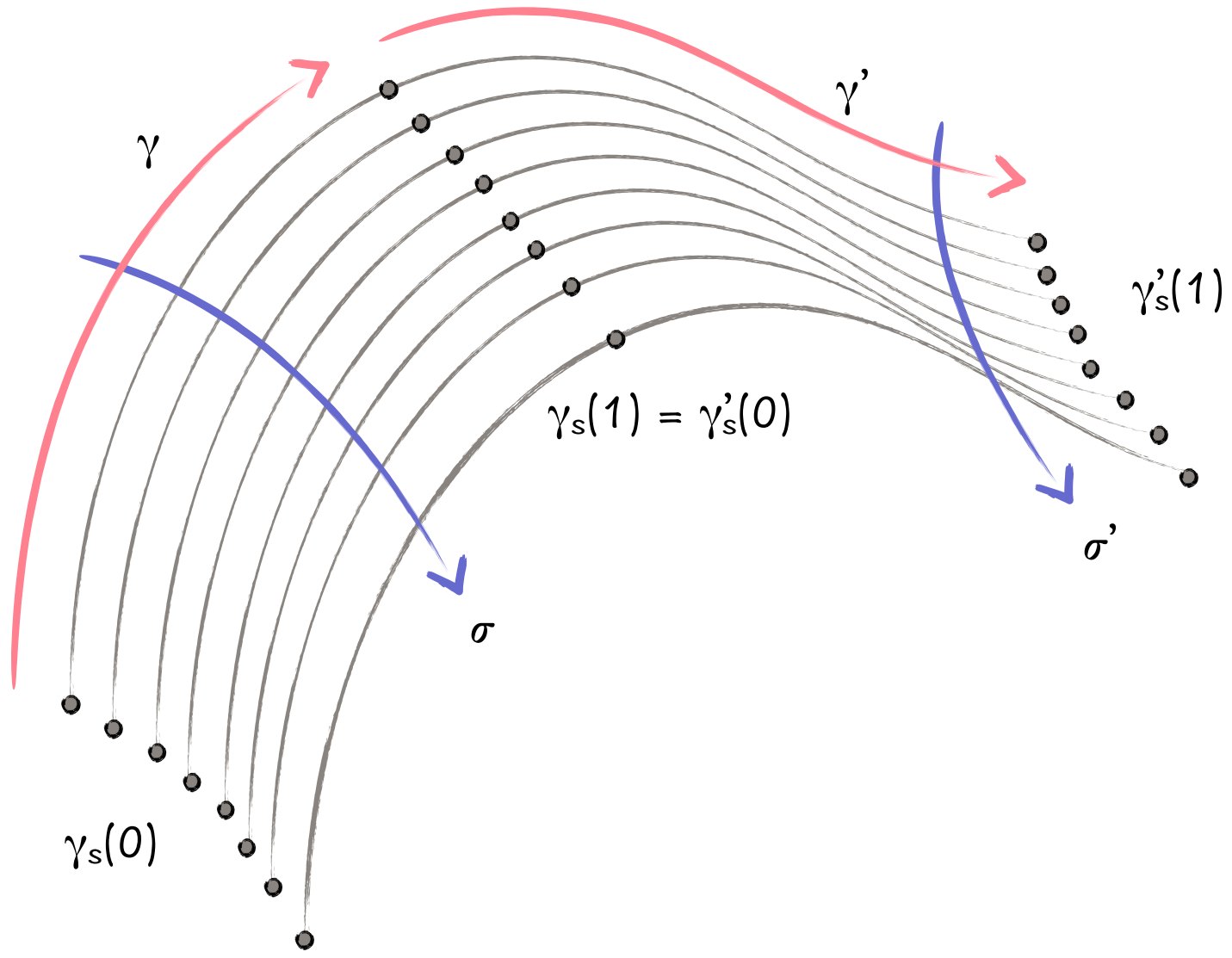

FIGURE 2. Integrating concatenations over homotopies.

and after a change of parameters $s' = 2s$ and $s'' = 2s - 1$, we get

$$\begin{aligned} \mathsf{K}\omega(\sigma * \sigma')_t(1) &= \int_0^1 \omega\left(\begin{pmatrix} s' \\ t \end{pmatrix} \mapsto \gamma_t(s')\right)_{\binom{s'}{t}} \begin{pmatrix} 1 \\ 0 \end{pmatrix}\begin{pmatrix} 0 \\ 1 \end{pmatrix} ds' \\ &+ \int_0^1 \omega\left(\begin{pmatrix} s'' \\ t \end{pmatrix} \mapsto \gamma'_t(s'')\right)_{\binom{s''}{t}} \begin{pmatrix} 1 \\ 0 \end{pmatrix}\begin{pmatrix} 0 \\ 1 \end{pmatrix} ds'' \\ &= \mathsf{K}\omega(\sigma)_t(1) + \mathsf{K}\omega(\sigma')_t(1). \end{aligned}$$

This is the first identity and we get the second by integration on both sides. □

**11. The Cocycle $\phi$.**

We consider the tautological pullback of ends : Paths(X) ⟶ X × X, that is, the space of pairs of paths in X with common endpoints:

$$\text{ends}^*(\text{Paths}(X)) = \{(\gamma, \gamma') \in \text{Paths}(X)^2 \mid \text{ends}(\gamma) = \text{ends}(\gamma')\},$$

equipped with the subset diffeology of the product diffeology.

(1) The map $\Phi : \text{ends}^*(\text{Paths}(X)) \to \text{Loops}(X)$, defined by

$$\Phi(\gamma, \gamma') = \gamma \vee \bar{\gamma}',$$

is a smooth homotopy equivalence. Then, since, by hypothesis, $\pi_1(X) = \{0\}$, the space Loops(X) is connected and also the space ends*(Paths(X)).

(2) The pullback of $\mathsf{K}\omega$ by $\Phi$ satisfies

$$\Phi^*(\mathsf{K}\omega) = \text{pr}_1^*(\mathsf{K}\omega) - \text{pr}_2^*(\mathsf{K}\omega).$$

It is also denoted by $\mathsf{K}\omega \ominus \mathsf{K}\omega$. This is a closed 1-form with the same periods as $\mathsf{K}\omega$ and an integration function given by

$$\phi = \Phi^*(f) = f \circ \Phi,$$

where $f$ is the integration function of $\mathsf{K}\omega \upharpoonright \mathrm{Loops}(\mathrm{X})$, that is,

$$\phi^*(\theta) = \Phi^*(\mathsf{K}\omega) = \mathsf{K}\omega \ominus \mathsf{K}\omega,$$

with $\theta$ the canonical 1-form on $\mathrm{T}_\omega$, defined uniquely by $\pi_\omega^*(\theta) = dt$, where $\pi_\omega$ is the canonical projection from $\mathbf{R}$ to $\mathrm{T}_\omega = \mathbf{R}/\mathrm{P}_\omega$.

(3) The smooth map $\phi$ is a (Chasles) cocycle, it can be chosen as

$$\phi(\gamma, \gamma') = \int_\gamma^{\gamma'} \mathsf{K}\omega_{x,x'} \mod \mathrm{P}_\omega, \qquad (\heartsuit)$$

where $\mathsf{K}\omega_{x,x'} = \mathsf{K}\omega \upharpoonright \mathrm{Paths}(\mathrm{X}, x, x')$ and $(x, x') = \mathrm{ends}(\gamma) = \mathrm{ends}(\gamma')$. The integral is computed along a path in this subspace of Paths(X). For any pairs $(\gamma, \gamma')$ and $(\gamma', \gamma'')$ of elements of ends*(Paths(X), one has:

$$\phi(\gamma, \gamma') + \phi(\gamma', \gamma'') = \phi(\gamma, \gamma'').$$

*Proof.* Let us recall that Paths(X) and Loops(X) are homotopy equivalent to stationary paths and loops, see [PIZ13, §5.5]. Then, we work with stationary paths and loops without recalling each time.

**Homotopy equivalence.** Let us begin with proving that ends*(Paths(X)) is homotopy equivalent to Loops(X). We consider the subspace $\mathrm{Loops}_{1/2}(\mathrm{X})$ of stationary loops that are stationary at $t = 1/2$, that is, constant on an open interval $]1/2-\varepsilon, 1/2+\varepsilon[$. The proof of [PIZ13, Ex. 84] can be adapted to show that Loops(X) and $\mathrm{Loops}_{1/2}(\mathrm{X})$ are homotopy equivalent. Now, let

$$\Phi : \mathrm{ends}^*(\mathrm{Paths}(\mathrm{X})) \to \mathrm{Loops}_{1/2}(\mathrm{X}), \ \text{ defined by } \ \Phi(\gamma, \gamma') = \gamma \vee \bar{\gamma}',$$

where $\bar{\gamma}'$ is the reverse path $\bar{\gamma}'(t) = \gamma'(1-t)$. Next, let

$$\bar{\Phi} : \mathrm{Loops}_{1/2}(\mathrm{X}) \to \mathrm{ends}^*(\mathrm{Paths}(\mathrm{X})), \ \text{ defined by } \ \bar{\Phi}(\ell) = (\gamma, \gamma')$$

with

$$\begin{cases} \gamma(t) = \ell(t/2), \ \text{ if } 0 \le t \le 1,\ \gamma(t) = \gamma(0) \text{ if } t \le 0 \text{ and } \gamma(t) = \ell(1/2) \text{ if } t \ge 1. \\ \gamma'(t) = \ell(1-t/2), \ \text{ if } 0 \le t \le 1,\ \gamma'(t) = \ell(1) \text{ if } t \le 0 \text{ and } \gamma'(t) = \ell(1/2) \text{ if } t \ge 1. \end{cases}$$

These two maps, $\Phi$ and $\bar{\Phi}$, are homotopic inverse to each other. Therefore, ends*(Paths(X)) is homotopy equivalent to Loops(X).

**Pullback $\mathsf{K}\omega$ by $\Phi$.** Now, since $\pi_1(\mathrm{X}, x) = \pi_0(\mathrm{Loops}(\mathrm{X}, x)) = \{0\}$, and $\pi_0(\mathrm{Loops}(\mathrm{X}))$ is a quotient of $\pi_0(\mathrm{Loops}(\mathrm{X}, x))$ [PIZ13, Ex. 87], it follows that Loops(X) is connected and so is ends*(Paths(X)). Moreover, the map $\Phi$ defined above satisfies the identity

$$\Phi^*(\mathsf{K}\omega) = \mathrm{pr}_1^*(\mathsf{K}\omega) - \mathrm{pr}_2^*(\mathsf{K}\omega).$$

Indeed, let $(\mathrm{P},\mathrm{P}')$ be a plot of ends*(Paths(X)). Then $\phi^*(\mathsf{K}\omega)(\mathrm{P},\mathrm{P}') = \mathsf{K}\omega(f\circ(\mathrm{P},\mathrm{P}')) = \mathsf{K}\omega(\mathrm{P}*\bar{\mathrm{P}}')$, where the operation $*$ has been defined in Article 10, and $\bar{\mathrm{P}}'(r)(t) = \mathrm{P}'(r)(1-t)$. Then, $\mathsf{K}\omega(\mathrm{P}*\bar{\mathrm{P}}') = \mathsf{K}\omega(\mathrm{P}) + \mathsf{K}\omega(\bar{\mathrm{P}}') = \mathsf{K}\omega(\mathrm{P}) - \mathsf{K}\omega(\mathrm{P}')$; that is, $\Phi^*(\mathsf{K}\omega) = \mathrm{pr}_1^*(\mathsf{K}\omega) - \mathrm{pr}_2^*(\mathsf{K}\omega)$. Therefore, $\mathsf{K}\omega\ominus\mathsf{K}\omega$ is closed and has the same periods as $\mathsf{K}\omega\restriction\mathrm{Loops(X)}$, as a consequence of the homotopic invariance of the de Rham cohomology [PIZ13, §6.88].

**Function $\phi$.** Next, we assumed that the periods $\mathrm{P}_\omega$ of the 1-form $\mathsf{K}\omega\restriction\mathrm{Loops(X)}$ are discrete.[10] Then, since Loops(X) is connected, let $f\in\mathrm{C}^\infty(\mathrm{Loops(X)},\mathrm{T}_\omega)$ be an integration function (Art. 9),

$$f\in\mathrm{C}^\infty(\mathrm{Loops(X)},\mathrm{T}_\omega) \text{ and } f^*(\theta) = \mathsf{K}\omega\restriction\mathrm{Loops(X)},$$

with $\theta$ the standard 1-form on $\mathrm{T}_\omega$. The function

$$\phi = f\circ\Phi\in\mathrm{C}^\infty(\mathrm{ends}^*(\mathrm{Paths(X)}),\mathrm{T}_\omega)$$

is an integration function on ends*(Paths(X)), i.e.:

$$\left[\mathrm{pr}_1^*(\mathsf{K}\omega) - \mathrm{pr}_2^*(\mathsf{K}\omega)\right] = \phi^*(\theta).$$

This function is explicitly given by

$$\phi(\gamma,\gamma') = \int_{\gamma'}^{\gamma}\mathsf{K}\omega\restriction\mathrm{Paths}(\mathrm{X},x,x') \mod \mathrm{P}_\omega, \qquad (\diamondsuit)$$

where the integral is computed along a path $\sigma$ in $\mathrm{Paths}(\mathrm{X},x,x')$, connecting $\gamma'$ to $\gamma$, where $(x,x') = \mathrm{ends}(\gamma) = \mathrm{ends}(\gamma')$. Indeed, by choice ot integration function $\phi = f\circ\Phi$, we have:

$$\phi(\gamma,\gamma') = \int_{\hat{x}}^{\gamma\vee\bar{\gamma}'}\mathsf{K}\omega\restriction\mathrm{Loops}(\mathrm{X},x) \mod \mathrm{P}_\omega,$$

where the integral is computed along a path in $\mathrm{Loops}(\mathrm{X},x)$, connecting $\hat{x}$ to $\gamma\vee\bar{\gamma}'$. Now, let us consider the map

$$\nu : \mathrm{Paths}(\mathrm{X},x,x')\to\mathrm{Loops}(\mathrm{X},x), \text{ with } \nu(\gamma') = \gamma\vee\bar{\gamma}'.$$

**Lemma.** One has: $\nu^*(\mathsf{K}\omega) = -\mathsf{K}\omega$.

*Proof of Lemma.* Remember that all paths are assumed to be stationary, then the concatenation is a smooth map [PIZ13, §5.4], and so $\nu$ is smooth. Now, since $\mathsf{K}\omega$ and $\nu^*(\mathsf{K}\omega)$ are 1-forms, we can compare $\nu^*(\mathsf{K}\omega)$ and $-\mathsf{K}\omega$ only on 1-plots [PIZ13, §6.37]. Let $s\mapsto\gamma'_s$ be a path in $\mathrm{Paths}(\mathrm{X},x,x')$. Then, $\nu^*(\mathsf{K}\omega)(s\mapsto\gamma'_s) = \mathsf{K}\omega(s\mapsto\gamma\vee\bar{\gamma}'_s)$. We have seen in Article 10 that $\mathsf{K}\omega([s\mapsto\gamma_s]\vee[s\mapsto\bar{\gamma}'_s]) = \mathsf{K}\omega(s\mapsto\gamma_s) + \mathsf{K}\omega(s\mapsto\bar{\gamma}'_s)$. Apply this identity to the constant plot $\gamma_s = \gamma$. The first term is then zero, and then: $\mathsf{K}\omega(s\mapsto\gamma\vee\bar{\gamma}'_s) = \mathsf{K}\omega(s\mapsto\bar{\gamma}'_s) = -\mathsf{K}\omega(s\mapsto\gamma'_s)$, from the change of integration parameter $t \mapsto 1-t$. Thus, $v^*(\mathsf{K}\omega)(s \mapsto \gamma'_s) = -\mathsf{K}\omega(s \mapsto \gamma'_s)$; that is, $v^*(\mathsf{K}\omega) = -\mathsf{K}\omega$. □

[10] Recall that in diffeology, 'discrete' means equipped with the discrete diffeology. In particular, any strict subgroup of **R** is discrete, even when it is dense.

Now, we use the variance of integration forms on chain [PIZ13, §6.67], that states, for $\sigma : s \mapsto \gamma'_s$, with $\text{ends}(\gamma'_s) = \text{ends}(\gamma)$ for all $s$ and $\gamma'_0 = \gamma$.

$$\int_{v_*(\sigma)} \mathsf{K}\omega = \int_\sigma v^*(\mathsf{K}\omega), \quad \text{which implies} \quad \int_{v_*(\sigma)} \mathsf{K}\omega = -\int_\sigma \mathsf{K}\omega\,.$$

But

$$\int_{v_*(\sigma)} \mathsf{K}\omega \;=\; \int_{s\mapsto\gamma\vee\bar\gamma'_s} \mathsf{K}\omega \;=\; \int_{\gamma\vee\bar\gamma}^{\gamma\vee\bar\gamma'} \mathsf{K}\omega \;=\; \underbrace{\int_{\gamma\vee\bar\gamma}^{\hat{x}} \mathsf{K}\omega}_{=0} + \int_{\hat{x}}^{\gamma\vee\bar\gamma'} \mathsf{K}\omega,$$

and, on the other hand

$$-\int_\sigma \mathsf{K}\omega \;=\; -\int_{s\mapsto\gamma'_s} \mathsf{K}\omega \;=\; -\int_\gamma^{\gamma'} \mathsf{K}\omega \;=\; \int_{\gamma'}^{\gamma} \mathsf{K}\omega\,.$$

Thus

$$\int_{\hat{x}}^{\gamma\vee\bar\gamma'} \mathsf{K}\omega \;=\; \int_{\gamma'}^{\gamma} \mathsf{K}\omega,$$

and this confirms the expression (♢) for $\phi$ above.

**Cocycle $\phi$.** The map $\phi$ defined by (♡) is clearly additive:

$$\begin{aligned}\phi(\gamma,\gamma') + \phi(\gamma',\gamma'') &= \int_\gamma^{\gamma'} \mathsf{K}\omega_{x,x'} \mod \mathrm{P}_\omega + \int_{\gamma'}^{\gamma''} \mathsf{K}\omega_{x,x'} \mod \mathrm{P}_\omega \\ &= \int_\gamma^{\gamma''} \mathsf{K}\omega_{x,x'} \mod \mathrm{P}_\omega = \phi(\gamma,\gamma'')\end{aligned}$$

for any triple of paths $\gamma$, $\gamma'$ and $\gamma''$ such that $\text{ends}(\gamma) = \text{ends}(\gamma') = \text{ends}(\gamma'') = (x,x')$. This completes the proof regarding the cocycle $\phi$. □

## 12. The Equivalence Relation $\gamma \sim_\omega \gamma'$.

The Chasles cocycle $\phi$ defined in the previous article leads to the definition of this equivalence relation: for any pair of paths $\gamma$ and $\gamma'$ in X:

$$\gamma \sim_\omega \gamma' \quad \text{iff} \quad \begin{cases} \text{ends}(\gamma) = \text{ends}(\gamma'), \\ \phi(\gamma,\gamma') = 0. \end{cases}$$

Let us make explicit the condition $\phi(\gamma,\gamma') = 0$: There exists a fixed-ends homotopy from $\gamma$ to $\gamma'$, inside $\text{Paths}(\mathrm{X}, x, x')$, with $(x,x') = \text{ends}(\gamma) = \text{ends}(\gamma')$, and:

$$\gamma \sim_\omega \gamma' \ \text{ iff } \ \text{ends}(\gamma) = \text{ends}(\gamma') \ \text{ and } \ \int_\gamma^{\gamma'} \mathsf{K}\omega_{x,x'} \in \mathrm{P}_\omega.$$

*Proof.* The fact that $\phi$ is a Chasles cocycle fulfills the requirements for $\sim_\omega$ to be an equivalence relation. □

**13. The Groupoid $\mathbf{T}_\omega$.**

The groupoid $\mathbf{T}_\omega$ defined by

$$\mathrm{Obj}(\mathbf{T}_\omega) = \mathrm{X} \text{ and } \mathrm{Mor}(\mathbf{T}_\omega) = \mathrm{Paths}(\mathrm{X})/\sim_\omega,$$

also denoted by $\mathscr{Y}$, is a diffeological fibrating groupoid:

(1) The composition $[\gamma]_\omega \cdot [\gamma']_\omega = [\gamma \vee \gamma']_\omega$ is well-defined and is associative.
(2) The inverse $[\gamma]_\omega^{-1} = [\bar\gamma]_\omega$ is well-defined.
(3) The identity $\mathbf{1}_x$ is $[\hat x : t \mapsto x]_\omega$, and the injection $j : x \mapsto \mathbf{1}_x$ from X to $\mathscr{Y}$ is an induction.
(4) The projection $\mathrm{ends} : \mathrm{Mor}(\mathbf{T}_\omega) \to \mathrm{X}\times\mathrm{X}$, defined by $\mathrm{ends}([\gamma]_\omega) = \mathrm{ends}(\gamma)$ is a subduction.

*Proof.* Let us prove that the groupoid operations are well defined

**Concatenation.** Let $\gamma_0$ and $\gamma'_0$ be two stationary paths such that $\gamma_0(1) = \gamma'_0(0)$. Then, let $\gamma_1$ and $\gamma'_1$ be two other paths such that $[\gamma_0]_\omega = [\gamma_1]_\omega$ and $[\gamma'_0]_\omega = [\gamma'_1]_\omega$. Because X is simply connected, there exists a fixed-ends homotopy $\sigma$ connecting $\gamma_0$ to $\gamma_1$, and another one $\sigma'$ connecting $\gamma'_0$ to $\gamma'_1$, that is, $\int_\sigma \mathsf{K}\omega \in \mathrm{P}_\omega$ and $\int_{\sigma'} \mathsf{K}\omega \in \mathrm{P}_\omega$. Thanks to Article 10, $\int_{\sigma*\sigma'} \mathsf{K}\omega = \int_\sigma \mathsf{K}\omega + \int_{\sigma'} \mathsf{K}\omega$, then $\int_{\sigma*\sigma'} \mathsf{K}\omega \in \mathrm{P}_\omega$, where $\sigma * \sigma'$ is a homotopy from $\gamma_0 \vee \gamma'_0$ to $\gamma_1 \vee \gamma'_1$. Thus $[\gamma_0 \vee \gamma'_0]_\omega = [\gamma_1 \vee \gamma'_1]_\omega$, and hence the composition $[\gamma]_\omega \cdot [\gamma']_\omega$ is well defined on $\mathscr{Y}$.

**Associativity, identities, and inverses.** The associativity of the concatenation, the fact that $\mathbf{1}_x = [\hat x]_\omega$ and $[\gamma]_\omega^{-1} = [\bar\gamma]_\omega$, where $\bar\gamma$ is the reverse of $\gamma$, are all based on the homotopies described in [PIZ13, §5.15 (Proof)], connecting $(\gamma_1 \vee \gamma_2) \vee \gamma_3$ to $\gamma_1 \vee (\gamma_2 \vee \gamma_3)$, $\gamma$ to $\hat x \vee \gamma$ and $\gamma \vee \bar\gamma$ to $\hat x$. The integral of $\mathsf{K}\omega$ over these homotopies vanishes. Indeed, let $\sigma : t \mapsto \gamma_t$ a path in Paths(X). The integrand of

$$\mathsf{K}\omega(\sigma)_t(1) = \int_0^1 \omega\left(\begin{pmatrix} s \\ t \end{pmatrix} \mapsto \gamma_t(s)\right)_{\binom{s}{t}} \begin{pmatrix} 1 \\ 0 \end{pmatrix}\begin{pmatrix} 0 \\ 1 \end{pmatrix} ds,$$

itself vanishes, because the homotopy $\sigma$ factorizes through a path (i.e., $\gamma_t(s) = \gamma'(\varphi(t,s))$ for some path $\gamma'$ and some real function $\varphi$), and the pullback of a 2-form on $\mathbf{R}$ vanishes.

**Groupoid $\mathbf{T}_\omega$.** Thus, $\mathbf{T}_\omega$ is a groupoid. Then, since the concatenation is smooth, as well as the inversion [PIZ13, §5.4 (2)], and since $x \mapsto [\hat x]_\omega$, where $\hat x$ is the constant path at $x$, is clearly an induction, $\mathbf{T}_\omega$ is a diffeological groupoid.

**$\mathbf{T}_\omega$ fibrating.** Since ends : Paths(X) $\to$ X $\times$ X is a subduction [PIZ13, §5.6] and $\mathrm{class}_\omega : \mathrm{Paths}(\mathrm{X}) \to \mathscr{Y}$ is smooth, the factorization $\mathrm{ends} : \mathscr{Y} \to \mathrm{X}\times\mathrm{X}$ is a subduction [PIZ13, §1.51]. Therefore, $\mathbf{T}_\omega$ is a fibrating groupoid [PIZ13, §8.4], and the $\mathscr{Y}_x$ are $\mathbf{T}_{\omega,x}$-principal fiber bundles attached to the points $x \in \mathrm{X}$ *loc. cit.*. □

## 14. The Prequantum 1-Form $\lambda$ and its Curvature.

By definition, the space of morphisms $\mathcal{Y}$ of the groupoid $\mathbf{T}_\omega$ is the quotient of Paths(X) by the equivalence relation $\sim_\omega$:

$$\gamma \sim_\omega \gamma' \quad \text{iff} \quad \begin{cases} \text{ends}(\gamma) = \text{ends}(\gamma'), \\ \phi(\gamma, \gamma') = 0, \end{cases}$$

where $\phi : \text{ends}^*(\text{Paths}(X)) \to T_\omega$ is the cocycle defined in Article 11. The condition $\phi(\gamma, \gamma') = 0$ is equivalent to

$$\gamma \sim_\omega \gamma' \ \text{iff} \ \int_{\gamma'}^{\gamma} \mathsf{K}\omega_{x,x'} \in P_\omega.$$

where $(x, x') = \text{ends}(\gamma) = \text{ends}(\gamma')$.

**Theorem.** *There exists a unique differential* 1*-form* $\lambda$ *on the space of morphisms* $\mathcal{Y} = \text{Mor}(\mathbf{T}_\omega)$ *such that* $\text{class}_\omega^*(\lambda) = \mathsf{K}\omega$*. We call* $\lambda$ *the* prequantum 1-form *on the groupoid* $\mathbf{T}_\omega$*. Furthermore,* $\lambda$ *is invariant under left and right composition in* $\mathbf{T}_\omega$*, and its curvature is given by* $d\lambda = \hat{1}^*\omega - \hat{0}^*\omega$*, where* $\hat{1}, \hat{0} : \mathcal{Y} \to X$ *are the target and source maps.*

*Proof.* **Existence and Uniqueness of $\lambda$.** We use the criterion for a differential form on a space W to be the pullback of a differential form on a quotient space $Z = W/\sim$ via the projection $p : W \to Z$. A differential form $\alpha \in \Omega^k(W)$ is the pullback of a unique differential form $\beta \in \Omega^k(Z)$ (i.e., $\alpha = p^*(\beta)$) if and only if $\text{pr}_1^*(\alpha) - \text{pr}_2^*(\alpha)$ vanishes on the total space of the tautological pullback $p^*(W) = \{(w_1, w_2) \in W \times W \mid p(w_1) = p(w_2)\}$, where $\text{pr}_1, \text{pr}_2 : p^*(W) \to W$ are the projection maps [PIZ13, §6.38 (Note 2)].

In our case, $W = \text{Paths}(X)$, $Z = \mathcal{Y}$, $p = \text{class}_\omega$, and $\alpha = \mathsf{K}\omega \in \Omega^1(\text{Paths}(X))$. The total space of the tautological pullback is $\text{class}_\omega^*(\text{Paths}(X)) = \{(\gamma, \gamma') \in \text{Paths}(X)^2 \mid \text{class}_\omega(\gamma) = \text{class}_\omega(\gamma')\}$. By definition of $\sim_\omega$, $\text{class}_\omega(\gamma) = \text{class}_\omega(\gamma')$ if and only if $\gamma \sim_\omega \gamma'$, which means $\text{ends}(\gamma) = \text{ends}(\gamma')$ and $\phi(\gamma, \gamma') = 0$. Thus, $\text{class}_\omega^*(\text{Paths}(X)) = \{(\gamma, \gamma') \in \text{ends}^*(\text{Paths}(X)) \mid \phi(\gamma, \gamma') = 0\}$.

We need to check if $\text{pr}_1^*(\mathsf{K}\omega) - \text{pr}_2^*(\mathsf{K}\omega)$ vanishes on this subspace. From Article 11, we know that $\text{pr}_1^*(\mathsf{K}\omega) - \text{pr}_2^*(\mathsf{K}\omega) = \phi^*(\theta)$ on $\text{ends}^*(\text{Paths}(X))$, where $\theta$ is the canonical 1-form on $T_\omega$. Thus:

$$(\text{pr}_1^*(\mathsf{K}\omega) - \text{pr}_2^*(\mathsf{K}\omega)) \restriction \text{class}_\omega^*(\text{Paths}(X)) = (\phi^*(\theta)) \restriction \phi^{-1}(0).$$

Since $\theta$ is a differential form on $T_\omega$, its pullback $\phi^*(\theta)$ restricted to the inverse image of a point, that is, $(\phi^*(\theta)) \restriction \phi^{-1}(0)$, is zero because differential forms vanish on constant plots [PIZ13, Ex. 96]. Indeed, let P be a plot in $\phi^{-1}(0)$, that is $\phi \circ P = 0$. Then, $\phi^*(\theta)(P) = \theta(\phi \circ P)$, but $\phi \circ P(r) = 0$, thus $\phi^*(\theta)(P) = 0$ for all P in $\phi^{-1}(0)$. The criterion is satisfied, which guarantees the existence and uniqueness of a differential 1-form $\lambda \in \Omega^1(\mathcal{Y})$ such that $\text{class}_\omega^*(\lambda) = \mathsf{K}\omega$.

**Invariance of $\boldsymbol{\lambda}$.** We need to show that for any $y_0 \in \mathscr{Y}$, the pullback of $\boldsymbol{\lambda}$ by left and right composition with $y_0$ is related to $\boldsymbol{\lambda}$. Let $y_0 = [\gamma_0]_\omega \in \mathscr{Y}$. The left composition map $\mathsf{L}(y_0) : \mathscr{Y}_{\mathrm{trg}(y_0),*} \to \mathscr{Y}_{\mathrm{src}(y_0),*}$ is given by $\mathsf{L}(y_0)(y) = y_0 \cdot y$. Similarly, the right composition map $\mathsf{R}(y_0) : \mathscr{Y}_{*,\mathrm{src}(y_0)} \to \mathscr{Y}_{*,\mathrm{trg}(y_0)}$ is given by $\mathsf{R}(y_0)(y) = y \cdot y_0$.

The invariance of $\mathsf{K}\omega$ under pre-concatenation and post-concatenation in Paths(X) is a known property [PIZ13, §6.85]. For any path $\gamma_0 \in \mathrm{Paths}(\mathrm{X})$, the map $\mathsf{L}(\gamma_0) : \mathrm{Paths}(\mathrm{X})_{\mathrm{trg}(\gamma_0),*} \to \mathrm{Paths}(\mathrm{X})_{\mathrm{src}(\gamma_0),*}$ given by $\mathsf{L}(\gamma_0)(\gamma) = \gamma_0 \vee \gamma$ satisfies $\mathsf{L}(\gamma_0)^*(\mathsf{K}\omega) = \mathsf{K}\omega$. Similarly, $\mathsf{R}(\gamma_0)^*(\mathsf{K}\omega) = \mathsf{K}\omega$.

The composition maps in $\mathscr{Y}$ are defined by factorization of path concatenation: $\mathsf{L}([\gamma_0]_\omega)([\gamma]_\omega) = [\gamma_0 \vee \gamma]_\omega = \mathrm{class}_\omega(\mathsf{L}(\gamma_0)(\gamma))$. This gives the commutative diagram:

$$\begin{array}{ccc}
\mathrm{Paths}(\mathrm{X})_{\mathrm{trg}(\gamma_0),*} & \xrightarrow{\ \mathsf{L}(\gamma_0)\ } & \mathrm{Paths}(\mathrm{X})_{\mathrm{src}(\gamma_0),*} \\
\mathrm{class}_\omega \big\downarrow & & \big\downarrow \mathrm{class}_\omega \\
\mathscr{Y}_{\mathrm{trg}(y_0),*} & \xrightarrow[\ \mathsf{L}(y_0)\ ]{} & \mathscr{Y}_{\mathrm{src}(y_0),*}
\end{array}$$

Taking the pullback by $\mathsf{L}(y_0)$ and using $\mathrm{class}_\omega^*(\boldsymbol{\lambda}) = \mathsf{K}\omega$:

$$\begin{aligned}
\mathsf{L}(y_0)^*(\boldsymbol{\lambda}) = \mathsf{L}(y_0)^*(\mathrm{class}_\omega^*)^{-1}(\mathsf{K}\omega) &= (\mathrm{class}_\omega \circ \mathsf{L}(\gamma_0))^*(\boldsymbol{\lambda}) \\
&= \mathsf{L}(\gamma_0)^*(\mathrm{class}_\omega^*(\boldsymbol{\lambda})) = \mathsf{L}(\gamma_0)^*(\mathsf{K}\omega).
\end{aligned}$$

By the invariance of $\mathsf{K}\omega$ under $\mathsf{L}(\gamma_0)$, $\mathsf{L}(\gamma_0)^*(\mathsf{K}\omega) = \mathsf{K}\omega$. Thus, $\mathsf{L}(y_0)^*(\boldsymbol{\lambda}) = \mathsf{K}\omega$. Since $\mathrm{class}_\omega^*(\boldsymbol{\lambda}) = \mathsf{K}\omega$, we have $\mathsf{L}(y_0)^*(\boldsymbol{\lambda}) = \mathrm{class}_\omega^*(\boldsymbol{\lambda})$. Restricting to the domain $\mathscr{Y}_{\mathrm{src}(y_0),*}$ and using the fact that $\mathrm{class}_\omega$ is a subduction from $\mathrm{Paths}(\mathrm{X})_{\mathrm{src}(\gamma_0),*}$ to $\mathscr{Y}_{\mathrm{src}(y_0),*}$, we can push the equality down to $\mathscr{Y}$:

$$\mathsf{L}(y_0)^*(\boldsymbol{\lambda}) \upharpoonright \mathscr{Y}_{\mathrm{src}(y_0),*} = \boldsymbol{\lambda} \upharpoonright \mathscr{Y}_{\mathrm{trg}(y_0),*}.$$

This shows left invariance. The proof for right invariance is analogous, using the postcomposition map $\mathsf{R}$.

**Curvature Formula.** With obvious notations: $\mathrm{ends}_{\mathscr{Y}} \circ \mathrm{class}_\omega = \mathrm{ends}_{\mathrm{Paths}}$. Then, Then, taking the pullback of $\mathrm{pr}_2^*(\omega) - \mathrm{pr}_1^*(\omega)$ by both sides of this equality, we get $\mathrm{class}_\omega^*(\mathrm{ends}_{\mathscr{Y}}^*(\mathrm{pr}_2^*(\omega) - \mathrm{pr}_1^*(\omega))) = \mathrm{ends}_{\mathrm{Paths}}^*(\mathrm{pr}_2^*(\omega) - \mathrm{pr}_1^*(\omega)) = \hat{1}_{\mathrm{paths}}^*(\omega) - \hat{0}_{\mathrm{paths}}^*(\omega) = d[\mathsf{K}\omega] = d[\mathrm{class}_\omega^*(\boldsymbol{\lambda})] = \mathrm{class}_\omega^*(d\boldsymbol{\lambda})$. Since $\mathrm{class}_\omega$ is a subduction, we get $d\boldsymbol{\lambda} = \mathrm{ends}_{\mathscr{Y}}^*(\mathrm{pr}_2^*\,\omega) - \mathrm{ends}_{\mathscr{Y}}^*(\mathrm{pr}_1^*\,\omega) = (\mathrm{pr}_2 \circ \mathrm{ends}_{\mathscr{Y}})^*\omega - (\mathrm{pr}_1 \circ \mathrm{ends}_{\mathscr{Y}})^*\omega = \hat{1}_{\mathscr{Y}}^*\,\omega - \hat{0}_{\mathscr{Y}}^*\,\omega$. which can also be written as $d\boldsymbol{\lambda} + \mathrm{ends}_{\mathscr{Y}}^*(\omega \ominus \omega) = 0$. □

## 15. The Isotropy $\mathbf{T}_{\omega,x}$.

By construction of the groupoid $\mathbf{T}_\omega$, the isotropy at a point $x \in \mathrm{X}$ is the subspace

$$\mathbf{T}_{\omega,x} := \mathrm{Mor}_{\mathbf{T}_\omega}(x,x) = \big\{[\ell]_\omega \mid \ell \in \mathrm{Loops}(\mathrm{X},x)\big\}.$$

Two loops $\ell$ and $\ell'$ based at $x$ are equivalent if and only if $\phi(\ell,\ell') = 0$. Define

$$\mathsf{F} : \mathrm{Loops}(\mathrm{X}, x) \to \mathrm{T}_\omega \ \text{ by } \ \mathsf{F}(\ell) = \pi_\omega\left(\int_{\hat{x}}^{\ell} \mathsf{K}\omega\right),$$

with $\hat{x}$ the constant loop $[t \mapsto x]$.

**Theorem.** *Two loops* $\ell, \ell' \in \mathrm{Loops}(\mathrm{X}, x)$ *are equivalent if and only if* $\mathsf{F}(\ell) = \mathsf{F}(\ell')$, *and the map* $\mathsf{F}$ *is a subduction. The projection* $\overline{\mathsf{F}} : \mathbf{T}_{\omega,x} \to \mathrm{T}_\omega$, *defined by* $\overline{\mathsf{F}}([\ell]_\omega) = \mathsf{F}(\ell)$ *is an isomorphism of diffeological groups.*

$$\begin{array}{ccc} \mathrm{Loops}(\mathrm{X}, x) & & \\ \Big\downarrow \scriptstyle{\mathrm{class}_\omega} & \searrow^{\mathsf{F}} & \\ \mathbf{T}_{\omega,x} = \mathrm{Loops}(\mathrm{X}, x)/\sim_\omega & \xrightarrow[\overline{\mathsf{F}}]{} & \mathrm{T}_\omega \end{array}$$

*Proof.* The notation $\mathbf{T}_{\omega,x}$ refers to the isotropy group of the groupoid $\mathbf{T}_\omega$ at the point $x \in \mathrm{X}$. That is, by definition:

$$\mathbf{T}_{\omega,x} = \{[\ell]_\omega \mid \ell \in \mathrm{Loops}(\mathrm{X}, x)\}.$$

Thus,

$$\mathbf{T}_{\omega,x} = \mathrm{Loops}(\mathrm{X}, x)/\sim_\omega, \ \text{ with } \ \ell \sim_\omega \ell' \ \text{ iff } \ \int_{\ell}^{\ell'} \mathsf{K}\omega_{x,x} \in \mathrm{P}_\omega.$$

Since $\pi_1(\mathrm{X}) = \{0\}$, $\ell$ and $\ell'$ are homotopic to the constant path $\hat{x} : t \mapsto x$. Moreover, by definition of $\mathrm{P}_\omega$,

$$\int_{\hat{x}}^{\ell} \mathsf{K}\omega_{x,x} + \int_{\ell}^{\ell'} \mathsf{K}\omega_{x,x} + \int_{\ell'}^{\hat{x}} \mathsf{K}\omega_{x,x} \in \mathrm{P}_\omega,$$

since this sum of integrals corresponds to the integral of $\mathsf{K}\omega_{x,x}$ on a loop in Loops(X). Hence

$$\begin{aligned} \int_{\ell}^{\ell'} \mathsf{K}\omega_{x,x} \in \mathrm{P}_\omega \ &\Leftrightarrow\ \int_{\hat{x}}^{\ell'} \mathsf{K}\omega_{x,x} - \int_{\hat{x}}^{\ell} \mathsf{K}\omega_{x,x} \in \mathrm{P}_\omega \\ &\Leftrightarrow\ \pi_\omega\left(\int_{\hat{x}}^{\ell'} \mathsf{K}\omega_{x,x}\right) = \pi_\omega\left(\int_{\hat{x}}^{\ell} \mathsf{K}\omega_{x,x}\right). \end{aligned}$$

Remember that $\pi_\omega : \mathbf{R} \to \mathrm{T}_\omega = \mathbf{R}/\mathrm{P}_\omega$. Hence

$$\ell \sim_\omega \ell' \ \Leftrightarrow\ \mathsf{F}(\ell) = \mathsf{F}(\ell') \ \text{ with } \ \mathsf{F}(\ell) = \pi_\omega\left(\int_{\hat{x}}^{\ell} \mathsf{K}\omega_{x,x}\right).$$

Hence, set theoretically, $\mathsf{F} : \mathrm{Loops}(\mathrm{X}, x) \to \mathrm{T}_\omega$ identifies $\mathrm{val}(\mathsf{F})$ with $\mathbf{T}_{\omega,x}$:

$$\mathbf{T}_{\omega,x} \equiv \mathrm{val}(\mathsf{F}).$$

**Lemma.** *The subset* $\mathrm{val}(\mathsf{F})$ *is a subgroup of* $\mathrm{T}_\omega$. *It can only be* $\{0\}$ *or* $\mathrm{T}_\omega$, *but if* $\omega \neq 0$, *then* $\mathrm{val}(\mathsf{F}) = \mathrm{T}_\omega$.

*Proof of Lemma.* Let $\mathrm{Paths}(\mathrm{Loops}(\mathrm{X}, x), \hat{x}, *)$ be the subspace of $\mathrm{Paths}(\mathrm{Loops}(\mathrm{X}))$ of homotopies connecting the constant loop $\hat{x}$ to any loop $\ell \in \mathrm{Loops}(\mathrm{X}, x)$. Define

$$\mathbf{F} : \mathrm{Paths}(\mathrm{Loops}(\mathrm{X}, x), \hat{x}, *) \to \mathbf{R}, \text{ by } \mathbf{F}(\sigma) = \int_\sigma \mathsf{K}\omega,$$

where $\sigma : s \mapsto \ell_s$ is a path in $\mathrm{Loops}(\mathrm{X}, x)$ connecting $\hat{x}$ to some loop $\ell$. Explicitly,

$$\mathbf{F}(\sigma) = \int_0^1 \mathsf{K}\omega(\sigma)_s(1) ds = \int_0^1 ds \int_0^1 \omega\left(\begin{pmatrix} t \\ s \end{pmatrix} \mapsto \ell_s(t)\right)_{\binom{t}{s}} \begin{pmatrix} 1 \\ 0 \end{pmatrix} \begin{pmatrix} 0 \\ 1 \end{pmatrix} dt$$

Let $\sigma$ and $\sigma'$ be two elements of $\mathrm{Paths}(\mathrm{Loops}(\mathrm{X}, x), \hat{x}, *)$, connecting $\hat{x}$ to $\ell$ and $\hat{x}$ to $\ell'$. Let $\bar{\sigma} : s \mapsto \bar{\ell}_s = [t \mapsto \ell_s(1-t)]$, hence $\bar{\sigma}$ is a path in $\mathrm{Loops}(\mathrm{X}, x)$ connecting $\hat{x}$ to the reverse $\bar{\ell}$. By a change of variable $t \mapsto 1-t$ in the expression of $\mathbf{F}(\sigma)$ above we have $\mathsf{K}\omega(\bar{\sigma}) = -\mathsf{K}\omega(\sigma)$. Also, since for all $s$ we have $\sigma(s)(1) = \sigma'(s)(0) = x$, we can consider the concatenation of homotopies $\sigma * \sigma'$. Then, thanks to Article 10, and to the property of $\bar{\sigma}$, we have:

$$\mathbf{F}(\sigma * \sigma') = \mathbf{F}(\sigma) + \mathbf{F}(\sigma') \text{ and } \mathbf{F}(\bar{\sigma}) = -\mathbf{F}(\sigma). \qquad (*)$$

Thus, if $t = \mathbf{F}(\sigma)$ and $t' = \mathbf{F}(\sigma')$, then $t + t' = \mathbf{F}(\sigma * \sigma') \in \mathrm{val}(\mathbf{F})$, and $-t = \mathbf{F}(\bar{\sigma}) \in \mathrm{val}(\mathbf{F})$. Therefore, $\mathrm{val}(\mathbf{F})$ is a subgroup of $\mathbf{R}$. But, $\mathrm{val}(\mathbf{F})$ is connected because the space of paths in $\mathrm{Loops}(\mathrm{X})$ based at $\hat{x}$, i.e., $\mathrm{Paths}(\mathrm{Loops}(\mathrm{X}, x), \hat{x}, *)$, is connected (actually contractible) and $\mathbf{F}$ is smooth. Thus, either $\mathrm{val}(\mathbf{F}) = \{0\}$ or $\mathrm{val}(\mathbf{F}) = \mathbf{R}$. Since $\mathrm{val}(\mathsf{F}) = \pi_\omega(\mathrm{val}(\mathbf{F}))$, then $\mathrm{val}(\mathsf{F}) = \{0\}$ or $\mathrm{val}(\mathsf{F}) = \mathrm{T}_\omega$. ☐

Let us examine the two different cases.

**The zero case.** If $\mathrm{val}(\mathsf{F}) = \{0\}$, then the isotropy $\mathbf{T}_{\omega,x}$ is reduced to the identity $\{1_x\}$, and $\mathscr{Y} = \mathrm{X} \times \mathrm{X}$. Thus, $\mathrm{d}\boldsymbol{\lambda} = \hat{1}^*(\omega) - \hat{0}^*(\omega)$. Restricted to $\mathscr{Y}_x = \{x\} \times \mathrm{X}$ we get $\omega = d\lambda$, and $\lambda = \boldsymbol{\lambda} \restriction \{x\} \times \mathrm{X}$, that is, a 1-form on X. But returning to $\mathsf{K}\omega$, we have $\mathsf{K}\omega = \mathsf{K}(d\lambda) = \hat{1}^*(\lambda) - \hat{0}^*(\lambda) - d(\mathsf{K}\lambda)$, then $\mathsf{K}\omega \restriction \mathrm{Loops}(\mathrm{X}, x) = d(\mathsf{K}\lambda)$ and the condition $\mathsf{F} = 0$ writes, for all $\ell \in \mathrm{Loops}(\mathrm{X}, x)$, that is:

$$0 = \int_{\hat{x}}^{\ell} \mathsf{K}\omega = \int_{\hat{x}}^{\ell} d[\mathsf{K}\lambda] = (\mathsf{K}\lambda)(\ell) - (\mathsf{K}\lambda)(\hat{x}) = \int_\ell \lambda - 0 = \int_\ell \lambda.$$

Indeed, by definition of the chain-homotopy operator $\mathsf{K}$, for a 1-form $\lambda$, $\mathsf{K}\lambda$ is the function from $\mathrm{Paths}(\mathrm{X})$ to $\mathbf{R}$:

$$(\mathsf{K}\lambda)(\gamma) = \int_0^1 \lambda(\gamma)_t(1)\, dt = \int_\gamma \lambda.$$

Therefore, $\lambda$ is a 1-form vanishing on every loop, thus $\lambda$ is closed and $\omega = 0$; see [PIZ13, Ex. 118], a case we excluded.

**The full case.** If $\operatorname{val}(\mathsf{F}) = \mathrm{T}_\omega$. Then, $\omega \neq 0$ and $\mathsf{F} : \operatorname{Loops}(\mathrm{X}, x) \to \mathrm{T}_\omega$ is a subduction projecting to a smooth isomorphism $\overline{\mathsf{F}} : \mathbf{T}_{\omega,x} \to \mathrm{T}_\omega$.

This is what we will prove now. The map $\mathsf{F}$ is already a smooth surjection, let us check that $\mathbf{F}$ is a subduction, which will imply that $\mathsf{F}$ itself is a subduction, and therefore that its projection $\overline{\mathsf{F}} : \mathbf{T}_{\omega,x} \to \mathrm{T}_\omega$ is a smooth isomorphism.

Let us choose a path

$$\sigma \in \operatorname{Paths}(\operatorname{Loops}(\mathrm{X}, x), \hat{x}, *) \ \text{ such that } \ \mathbf{F}(\sigma) \neq 0.$$

Such path exists since $\mathbf{F} : \operatorname{Paths}(\operatorname{Loops}(\mathrm{X}, x), \hat{x}, *) \to \mathbf{R}$ is surjective. Let

$$\sigma_s(t) = \sigma(st) \ \text{ and } \ \varphi(s) = \mathbf{F}(\sigma_s).$$

So, $\varphi$ is a smooth real function such that

$$\varphi(0) = 0 \ \text{ and } \ \varphi(1) = \mathbf{F}(\sigma) \neq 0,$$

since $\varphi(0) = \mathbf{F}(\sigma_0)$ and $\sigma_0 = [t \mapsto \sigma(0) = \hat{x}]$. Thus, there exists

$$s_0 \in \,]0,1[ \ \text{ such that } \ \varphi'(s_0) \neq 0,$$

where $\varphi'$ denotes the derivative of $\varphi$. Otherwise $\varphi$ would be constant and then equal to 0. Let

$$\mathfrak{S}(s) = \bar{\sigma}_{s_0} * \sigma_{s+s_0}, \ \text{ where } \ \bar{\sigma}_{s_0}(t) = [t' \mapsto \sigma_{s_0}(t)(1-t')].$$

Define

$$\psi = \mathbf{F} \circ \mathfrak{S}.$$

That is, $\psi \in \mathrm{C}^\infty(\mathbf{R}, \mathbf{R})$ with:

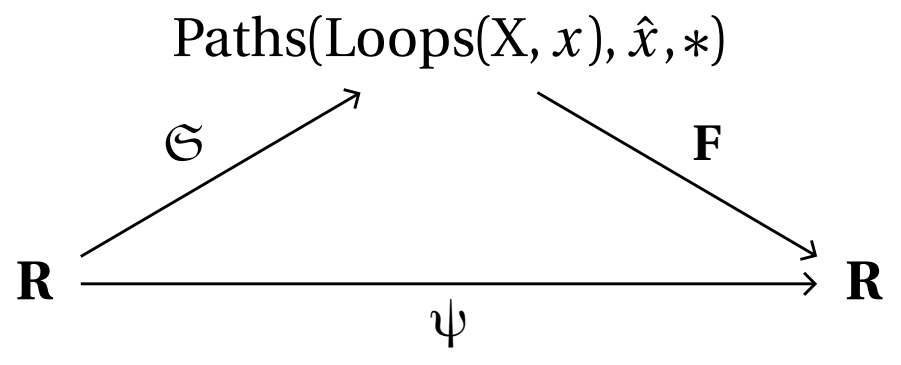

By Article 10, we have:

$$\mathbf{F}(\bar{\sigma}_{s_0} * \sigma_{s+s_0}) = \mathbf{F}(\sigma_{s+s_0}) - \mathbf{F}(\sigma_{s_0}),$$

and then

$$\psi(s) = \mathbf{F}(\mathfrak{S}(s)) = \mathbf{F}(\bar{\sigma}_{s_0} * \sigma_{s+s_0}) = \mathbf{F}(\sigma_{s+s_0}) - \mathbf{F}(\sigma_{s_0}) = \varphi(s+s_0) - \varphi(s_0).$$

Thus,

$$\psi(0) = 0 \ \text{ and } \ \psi'(0) = \varphi'(s_0) \neq 0.$$

Therefore, by the inverse function theorem on $\mathbf{R}$, there exists $\varepsilon > 0$ such that $\psi \restriction \,]-\varepsilon, +\varepsilon[$ is a local diffeomorphism mapping $]-\varepsilon, +\varepsilon[$ to some open interval $]-a, +b[$ with $a, b > 0$ and $\psi^{-1}(0) = 0$. Hence,

$$\psi \restriction \,]-\varepsilon, +\varepsilon[ \,= \mathbf{F} \circ \mathfrak{S} \restriction \,]-\varepsilon, +\varepsilon[ \ \text{ implies } \ \psi \circ \psi^{-1} = \mathbf{1}_{]-\varepsilon,+\varepsilon[} = \mathbf{F} \circ (\mathfrak{S} \circ \psi^{-1}).$$

Therefore,

$$\mathbf{F} \circ (\mathfrak{S} \circ \psi^{-1}) = \mathbf{1}_{]-\varepsilon,+\varepsilon[}.$$

Let us define:

$$\Sigma = \mathfrak{S} \circ \psi^{-1},$$

Hence,

$$\begin{cases} \Sigma : ]-a,+b[ \to \mathrm{Paths}(\mathrm{Loops}(\mathrm{X},x),\hat{x},*), \\ \text{with } \Sigma(0) = \mathfrak{S}(0) = \bar{\sigma}_{s_0} \vee \sigma_{s_0}. \end{cases}$$

is a local smooth section of $\mathbf{F} : \mathrm{Paths}(\mathrm{Loops}(\mathrm{X},x),\hat{x},*) \to \mathbf{R}$.

$$\mathbf{F} \circ \Sigma = \mathbf{1}_{]-\varepsilon,+\varepsilon[},$$

defined on an open interval around $0 \in \mathbf{R}$ and mapping 0 to $\mathfrak{S}(0) = \bar{\sigma}_{s_0} \vee \sigma_{s_0}$, and for any plot P in $\mathbf{R}$ mapping $r_0$ to 0, $\mathrm{Q} = \Sigma \circ \mathrm{P}$ is a lift in $\mathrm{Paths}(\mathrm{Loops}(\mathrm{X},x),\hat{x},*)$:

$$\begin{array}{ccc} \mathrm{Paths}(\mathrm{Loops}(\mathrm{X},x),\hat{x},*) & & & \mathrm{Paths}(\mathrm{Loops}(\mathrm{X},x),\hat{x},*) \\ \Sigma = \mathfrak{S}\circ\psi^{-1} \uparrow \downarrow \mathbf{F} & \text{implies} & \overset{\mathrm{Q}}{\nearrow} & \Sigma \uparrow \downarrow \mathbf{F} \\ ]-\varepsilon,+\varepsilon[ \subset \mathbf{R} & & \mathrm{U} \xrightarrow[\mathrm{P}]{} & ]-\varepsilon,+\varepsilon[ \subset \mathbf{R} \end{array}$$

Now, thanks to the additivity of function $\mathbf{F}$, we can translate this local section everywhere on $\mathbf{R}$. Let $t_0 \in \mathbf{R}$, and recall that $\mathbf{F}$ is surjective. Thus, there exists $\sigma' \in \mathrm{Paths}(\mathrm{Loops}(\mathrm{X},x),\hat{x},*)$ such that $\mathbf{F}(\sigma') = t_0$. Let us then consider

$$\mathfrak{S}'(s) = \sigma' * \mathfrak{S}(s),$$

and let,

$$\Psi(s) = \mathbf{F}(\mathfrak{S}'(s)), \text{ that is, } \mathbf{F}(\sigma' * \mathfrak{S}(s)) = \mathbf{F}(\sigma') + \mathbf{F}(\mathfrak{S}(s)) = t_0 + \psi(s).$$

So,

$$\Psi(0) = t_0 \text{ and } \Psi'(0) = \psi'(0) \neq 0.$$

Thus, $\Psi$ is a local diffeomorphism of $\mathbf{R}$ mapping 0 to $t_0$, its inverse $\Psi^{-1}$ maps $t_0$ to 0. We define then:

$$\Sigma' = \mathfrak{S}' \circ \Psi^{-1}$$

Then, $\Sigma'$ is a local smooth section of $\mathbf{F}$ over the interval $]t_0 - \varepsilon, t_0 + \varepsilon[$,

$$\mathbf{F} \circ \Sigma' = \mathbf{1}_{]t_0-\varepsilon,t_0+\varepsilon[}, \text{ such that } \Sigma'(t_0) = \sigma' * \mathfrak{S}(0) = \sigma' * (\bar{\sigma}_{s_0} \vee \sigma_{s_0}).$$

Hence, we can lift locally $\mathbf{F}$ around every point $t_0 \in \mathbf{R}$. Therefore $\mathbf{F}$ is a subduction. Then, by projection, $\mathsf{F} = \pi_\omega \circ \mathbf{F}$ is itself a subduction, and the mapd $\bar{\mathsf{F}} : \mathbf{T}_{\omega,x} \to \mathrm{T}_\omega$, which is injective by construction, is a diffeomorphism and therefore a smooth isomorphism from $\mathbf{T}_{\omega,x}$ to $\mathrm{T}_\omega$. In conclusion, $\mathbf{T}_\omega$ is a fibrating groupoid with isotropy $\mathrm{T}_\omega$. ☐

## IV. Symmetries of the Prequantum Structure

In this section, we investigate the symmetries of the prequantum groupoid $(\mathbf{T}_\omega, \boldsymbol{\lambda})$ constructed in the Main Theorem. We show that the group of automorphisms of this prequantum structure is naturally isomorphic to the group of $\omega$-preserving diffeomorphisms of the base space X, demonstrating a faithful representation of the classical symmetries at the prequantum level.

### 16. Automorphisms Induced by Symmetries.

This article discusses the action of symmetries. The group of diffeomorphisms of X that preserve the closed 2-form $\omega$, denoted by $\mathrm{Diff}(\mathrm{X}, \omega)$, plays a crucial role in the symmetry analysis of the system $(\mathrm{X}, \omega)$. In this framework, these symmetries lift naturally to automorphisms of the prequantum groupoid $(\mathbf{T}_\omega, \boldsymbol{\lambda})$. Let us denote

$$\mathrm{Aut}(\mathbf{T}_\omega, \boldsymbol{\lambda}) = \{(\Phi, \phi) \in \mathrm{Aut}(\mathbf{T}_\omega) \mid \Phi^*(\boldsymbol{\lambda}) = \boldsymbol{\lambda}\}.$$

On the other hand, the group of automorphisms of $(\mathrm{X}, \omega)$ is denoted by

$$\mathrm{Diff}(\mathrm{X}, \omega) = \{\phi \in \mathrm{Diff}(\mathrm{X}) \mid \phi^*(\omega) = \omega\}.$$

Note first that:

**Proposition 1.** *If* $(\Phi, \phi) \in \mathrm{Aut}(\mathbf{T}_\omega, \boldsymbol{\lambda})$ *then* $\phi \in \mathrm{Diff}(\mathrm{X}, \omega)$.

And then, the construction of automorphisms of the prequantum groupoid:

**Proposition 2..** *Let* $\mathbf{T}_\omega$ *be the associated prequantum groupoid and* $\boldsymbol{\lambda}$ *its prequantum* 1*-form, associated with the system* $(\mathrm{X}, \omega)$*, where* X *is connected and simply connected.*

*For any* $\phi \in \mathrm{Diff}(\mathrm{X}, \omega)$*, that is,* $\phi \in \mathrm{Diff}(\mathrm{X})$ *and* $\phi^*(\omega) = \omega$*, define the map*

$$\Phi : \mathscr{Y} \to \mathscr{Y} \quad \textit{with} \quad \Phi([\gamma]_\omega) = [\phi \circ \gamma]_\omega.$$

*Then, this automorphism preserves the prequantum* 1*-form* $\boldsymbol{\lambda}$, $\Phi^*\boldsymbol{\lambda} = \boldsymbol{\lambda}$*, and*

$$(\Phi, \phi) \in \mathrm{Aut}(\mathbf{T}_\omega, \boldsymbol{\lambda}).$$

Note that this proposition shows that the map $(\Phi, \phi) \mapsto \phi$ from $\mathrm{Aut}(\mathbf{T}_\omega, \boldsymbol{\lambda})$ to $\mathrm{Diff}(\mathrm{X}, \omega)$ is surjective.

*Proof.* We first show that $\Phi$ is well-defined. Suppose $\gamma \sim_\omega \gamma'$. By definition, let $(x, x') = \mathrm{ends}(\gamma) = \mathrm{ends}(\gamma')$. Then, for any fixed-ends homotopy $\sigma : s \mapsto \gamma_s$ in $\mathrm{ends}^{-1}(x, x')$, from $\gamma = \gamma_0$ to $\gamma' = \gamma_1$,

$$\int_\sigma \mathsf{K}\omega \in \mathrm{P}_\omega.$$

Consider the path $\phi \circ \sigma : \mathbf{R} \to \mathrm{Paths}(\mathrm{X})$ defined by $s \mapsto \phi \circ \gamma_s$. This is a fixed-ends homotopy from $\phi \circ \gamma$ to $\phi \circ \gamma'$, with $\phi \circ \gamma_s(0) = \phi(x)$ and $\phi \circ \gamma_s(1) = \phi(x')$. That is, $\phi \circ \sigma$ is a path in $\mathrm{ends}^{-1}(\phi(x), \phi(x'))$. Therefore,

$$\int_{\phi\circ\sigma} \mathsf{K}\omega = \int_{\sigma} \phi^*(\mathsf{K}\omega) = \int_{\sigma} \mathsf{K}\omega \in \mathrm{P}_\omega,$$

thanks to the variance of the chain-homotopy operator [PIZ13, §6.84]. Thus, $\Phi$ is well-defined.

Next, $\Phi$ is a smooth map from $\mathscr{Y}$ to $\mathscr{Y}$. Indeed, the map $\Psi : \mathrm{Paths(X)} \to \mathrm{Paths(X)}$ defined by $\Psi(\gamma) = \phi \circ \gamma$ is smooth, as it is the composition operator in the functional diffeology. Since $\mathrm{class}_\omega : \mathrm{Paths(X)} \to \mathscr{Y}$ is a subduction and $\Phi \circ \mathrm{class}_\omega = \mathrm{class}_\omega \circ \Psi$, $\Phi$ is smooth because $\Psi$ is smooth.

The map $\Phi$ is bijective because $\phi$ is a diffeomorphism. Its inverse is given by $\Phi^{-1}([\gamma']_\omega) = [\phi^{-1} \circ \gamma']_\omega$.

Finally, we verify the functor properties:

**(a)** Preservation of Source and Target: For $[\gamma]_\omega \in \mathscr{Y}$,

$$\begin{aligned}\mathrm{ends} \circ \Phi([\gamma]_\omega) &= \mathrm{ends}([\phi \circ \gamma]_\omega) = \mathrm{ends}(\phi \circ \gamma)\\ &= (\phi(\gamma(0)), \phi(\gamma(1))) = (\phi \times \phi)(\mathrm{ends}[\gamma]_\omega).\end{aligned}$$

**(b)** Preservation of Composition: For composable $[\gamma]_\omega, [\gamma']_\omega \in \mathscr{Y}$,

$$\begin{aligned}\Phi([\gamma]_\omega \cdot [\gamma']_\omega) &= \Phi([\gamma \vee \gamma']_\omega) = [\phi \circ (\gamma \vee \gamma')]_\omega\\ &= [(\phi \circ \gamma) \vee (\phi \circ \gamma')]_\omega = [\phi \circ \gamma]_\omega \cdot [\phi \circ \gamma']_\omega = \Phi([\gamma]_\omega) \cdot \Phi([\gamma']_\omega).\end{aligned}$$

**(c)** Preservation of Identity: For $x \in \mathrm{X}$, with $\hat{x} = [t \mapsto x]$,

$$\Phi(\mathbf{1}_x) = \Phi([\hat{x}]_\omega) = [\phi \circ \hat{x}]_\omega = [\widehat{\phi(x)}]_\omega = \mathbf{1}_{\phi(x)}.$$

Since $(\phi, \Phi)$ is a groupoid morphism and both $\phi$ and $\Phi$ are diffeomorphisms, $(\phi, \Phi)$ is a diffeological groupoid automorphism of $\mathbf{T}_\omega$. This proves part (1).

For part (2), we show that $\Phi^*\boldsymbol{\lambda} = \boldsymbol{\lambda}$. Let $\phi_* : \mathrm{Paths(X)} \to \mathrm{Paths(X)}$ defined by $\phi_*(\gamma) = \phi \circ \gamma$. The map $\Phi$ is defined by the commutative diagram:

$$\begin{array}{ccc}\mathrm{Paths(X)} & \xrightarrow{\phi_*} & \mathrm{Paths(X)}\\ \mathrm{class}_\omega \downarrow & & \downarrow \mathrm{class}_\omega\\ \mathscr{Y} & \xrightarrow[\Phi]{} & \mathscr{Y}\end{array}$$

Then, $(\Phi \circ \mathrm{class}_\omega)^*(\boldsymbol{\lambda}) = (\mathrm{class}_\omega \circ \phi_*)^*(\boldsymbol{\lambda})$. That is, $\mathrm{class}_\omega^*(\Phi^*(\boldsymbol{\lambda})) = (\phi_*)^*(\mathrm{class}_\omega^*(\boldsymbol{\lambda}))$. Thus, $\mathrm{class}_\omega^*(\Phi^*(\boldsymbol{\lambda})) = (\phi_*)^*(\mathsf{K}\omega)$. But, since $\phi^*(\omega) = \omega$, thanks to the variance of the chain-homotopy operator $\mathsf{K}$, see [PIZ13, §6.84], $(\phi_*)^*(\mathsf{K}\omega) = \mathsf{K}\omega$, hence: $\mathrm{class}_\omega^*(\Phi^*(\boldsymbol{\lambda})) = \mathsf{K}\omega = \mathrm{class}_\omega(\boldsymbol{\lambda})$. Thus, since $\mathrm{class}_\omega$ is a subduction, thanks to [PIZ13, §6.39], $\Phi^*(\boldsymbol{\lambda}) = \boldsymbol{\lambda}$. ☐

## 17. The Automorphism Group of $(\mathbf{T}_\omega, \lambda)$.

In this article, we discuss the automorphism group of the structure $(\mathbf{T}_\omega, \boldsymbol{\lambda})$ and show that its group of automorphisms is isomorphic to the group of symmetries of the structure $(\mathrm{X}, \omega)$, i.e., $\mathrm{Diff}(\mathrm{X}, \omega)$.

The group of automorphisms of the groupoid $\mathbf{T}_\omega$ preserving the prequantum 1-form $\boldsymbol{\lambda}$ is denoted by $\mathrm{Aut}(\mathbf{T}_\omega, \boldsymbol{\lambda})$. An element $(\Phi, \phi) \in \mathrm{Aut}(\mathbf{T}_\omega, \boldsymbol{\lambda})$ is an element $(\Phi, \phi) \in \mathrm{Aut}(\mathbf{T}_\omega)$ such that $\Phi^*(\boldsymbol{\lambda}) = \boldsymbol{\lambda}$. That is:

(1) $\Phi \in \mathrm{Diff}(\mathscr{Y})$, with $\mathscr{Y} = \mathrm{Mor}(\mathbf{T}_\omega)$.
(2) There exists $\phi \in \mathrm{Diff}(\mathrm{X})$ such that: $\mathrm{ends} \circ \Phi = (\phi \times \phi) \circ \mathrm{ends}$,

$$\begin{array}{ccc} \mathscr{Y} & \xrightarrow{\Phi} & \mathscr{Y} \\ {\scriptstyle \mathrm{ends}}\downarrow & & \downarrow{\scriptstyle \mathrm{ends}} \\ \mathrm{X}\times\mathrm{X} & \xrightarrow[\phi\times\phi]{} & \mathrm{X}\times\mathrm{X} \end{array} \qquad \text{with} \quad (\phi \times \phi)(x, x') = (\phi(x), \phi(x')).$$

(3) $\Phi(y \cdot y') = \Phi(y) \cdot \Phi(y')$, for all juxtaposable pairs.
(4) $\Phi^*(\boldsymbol{\lambda}) = \boldsymbol{\lambda}$.

**Proposition.** *Let* $(\Phi, \phi)$ *be a pair in* $\mathrm{Aut}(\mathbf{T}_\omega)$. *If* $\Phi^*(\boldsymbol{\lambda}) = \boldsymbol{\lambda}$, *then* $\phi \in \mathrm{Diff}(\mathrm{X}, \omega)$.

We finally establish the most important result on the group of automorphisms of the prequantum groupoid:

**Theorem.** *The projection* $(\Phi, \phi) \mapsto \phi$ *from the group of automorphisms of the prequantum groupoid* $\mathrm{Aut}(\mathbf{T}_\omega, \boldsymbol{\lambda})$ *to the group of symmetries of the parasymplectic form* $\omega$ *on* $\mathrm{X}$, $\mathrm{Diff}(\mathrm{X}, \omega)$, *is an isomorphism of diffeological groups:*

$$\mathrm{Aut}(\mathbf{T}_\omega, \boldsymbol{\lambda}) \simeq \mathrm{Diff}(\mathrm{X}, \omega).$$

**Note.** This theorem reveals a particularly interesting situation: the group of $\omega$-preserving diffeomorphisms of the base space $(\mathrm{X}, \omega)$ has a full and faithful representation as automorphisms of the prequantum groupoid $(\mathbf{T}_\omega, \boldsymbol{\lambda})$. This is a significant feature, as it implies that every symmetry of the classical system $(\mathrm{X}, \omega)$ lifts to a unique automorphism of the prequantum structure, and the entire group $\mathrm{Diff}(\mathrm{X}, \omega)$ is represented at the prequantum level without any loss of information or the need for a central extension. The prequantum property of the system $(\mathrm{X}, \omega)$ is captured in the isotropy groups $\mathbf{T}_{\omega,x}$ of the groupoid $\mathbf{T}_\omega$. This is achieved without abandoning any symmetries of the structure.

*Proof.* Let us begin by proving the first assertion. Let us apply $\mathrm{ends} \circ \Phi = (\phi \times \phi) \circ \mathrm{ends}$ to the pullback of $\omega \ominus \omega$ :

$$\begin{aligned} (\mathrm{ends} \circ \Phi)^*(\omega \ominus \omega) &= ((\phi \times \phi) \circ \mathrm{ends})^*(\omega \ominus \omega) \\ \Phi^*(\mathrm{ends}^*(\omega \ominus \omega)) &= \mathrm{ends}^*((\phi \times \phi)^*(\omega \ominus \omega)) \\ \Phi^*(\mathrm{d}\boldsymbol{\lambda}) &= \mathrm{ends}^*(\phi^*(\omega) \ominus \phi^*(\omega)) \end{aligned}$$

$$\begin{aligned} d[\Phi^*\boldsymbol{\lambda}] &= \mathrm{ends}^*(\phi^*(\omega)\ominus\phi^*(\omega)) \\ d\boldsymbol{\lambda} &= \mathrm{ends}^*(\phi^*(\omega)\ominus\phi^*(\omega)) \\ \mathrm{ends}^*(\omega\ominus\omega) &= \mathrm{ends}^*(\phi^*(\omega)\ominus\phi^*(\omega)) \end{aligned}$$

Since ends is a subduction, this implies $\phi^*(\omega)\ominus\phi^*(\omega) = \omega\ominus\omega$, see [PIZ13, §6.39]. Now apply that to a plot $\mathrm{P}\times\mathrm{Const}$, where Const is any constant plot. Then we get $\phi^*(\omega)(\mathrm{P}) = \omega(\mathrm{P})$ for all plots in X, i.e., $\phi^*(\omega) = \omega$.

Now, let us consider the projection $\mathrm{pr}_2(\Phi,\phi)\mapsto\phi$ from $\mathrm{Aut}(\mathbf{T}_\omega,\boldsymbol{\lambda})$ to $\mathrm{Diff}(\mathrm{X},\omega)$. We know already by Article 16 that this is a surjection. Let us consider the kernel $\ker(\mathrm{pr}_2)$ of this projection, that is the subgroup of $(\Phi,\mathbf{1}_\mathrm{X})\in\mathrm{Aut}(\mathbf{T}_\omega,\boldsymbol{\lambda})$. Since $\Phi$ projects onto the identity of X, one has:

$$\mathrm{ends}\circ\Phi = \mathrm{ends}.$$

That is, for all $y\in\mathscr{Y}$, $\mathrm{ends}(\Phi(y)) = \mathrm{ends}(y)$. Thus, $\Phi(y)$ is composable with $y^{-1}$, and $\tau(y) = \Phi(y)\cdot y^{-1}$ belongs to the isotropy group $\mathbf{T}_{\omega,x}$, with $x = \mathrm{src}(y)$. By construction, the map $\tau$ is smooth. Therefore the map $\Phi\in\ker(\mathrm{pr}_2)$ writes $\Phi(y) = \tau(y)\cdot y$, and satisfies $\Phi^*(\boldsymbol{\lambda}) = \boldsymbol{\lambda}$.

Now choose $x\in\mathrm{X}$, and let $\hat{1}_x:\mathrm{Y}_x\to\mathrm{X}$, where $\mathrm{Y}_x = \mathrm{Mor}_{\mathbf{T}_\omega}(x,*)$ and $\hat{1}_x = \hat{1}$. Let $\lambda_x = \boldsymbol{\lambda}\restriction\mathrm{Y}_x$. The projection $\hat{1}_x$ is a principal fibration with group $\mathrm{T}_x = \mathbf{T}_{\omega,x}$, and the 1-form $\lambda_x$ is actually a diffeological connection form satisfying the conditions [PIZ13, §8.37]:

* The form is *invariant* by $\mathrm{T}_x$, $\tau^*(\lambda_x) = \lambda_x$, for all $\tau\in\mathrm{T}_x$.
* The form is *calibrated*: $\hat{y}^*(\lambda_x) = \theta$, where $\hat{y}:\tau\mapsto\tau(y)$ is the *orbit map*, and $\theta$ is the canonical 1-form on $\mathrm{T}_x\simeq\mathbf{R}/\mathrm{P}_\omega$.

Now, the restriction $\Phi_x$ of $\Phi$ on $\mathrm{Y}_x$ writes the same way, that is, $\Phi_x(y) = \tau(y)\cdot y$.

**Lemma.** *The automorphism* $\Phi_x : y\mapsto\tau(y)\cdot y$ *preserves* $\lambda_x$, *i.e.,* $\Phi_x^*(\lambda_x) = \lambda_x$, *if and only if* $\tau(y) = \tau_x$ *is constant on* $\mathrm{Y}_x$.

*Proof of the lemma.* Let us compute $\Phi_x^*(\lambda_x)$ on a plot $r\mapsto y_r$. That is,

$$\Phi_x^*(\lambda_x)(r\mapsto y_r)_r(\delta r) = \lambda_x(r\mapsto\tau(y_r)\cdot y_r)_r(\delta r),$$

for all $r$ in the domain of the plot and $\delta r$ a tangent vector at the point $r$. For that, we will separate the variables: for any integers $n$ and $m$, any $n$-plot $\boldsymbol{\tau}:\mathrm{U}\to\mathrm{T}_x$ and any $m$-plot $\mathrm{P}:\mathrm{V}\to\mathrm{Y}_x$, let $\boldsymbol{\tau}\cdot\mathrm{P}$ be the plot of $\mathrm{Y}_x$ defined by

$$\boldsymbol{\tau}\cdot\mathrm{P}:\mathrm{U}\times\mathrm{V}\to\mathrm{Y}_x, \text{ with } \boldsymbol{\tau}\cdot\mathrm{P}:(r,s)\mapsto\boldsymbol{\tau}(r)\cdot\mathrm{P}(s).$$

Let $(r,s)\in\mathrm{U}\times\mathrm{V}$ and two vectors $\delta r\in\mathbf{R}^n$ and $\delta s\in\mathbf{R}^m$. The value of $\lambda_x$ on $\boldsymbol{\tau}\cdot\mathrm{P}$ is given by

$$\lambda_x(\boldsymbol{\tau}\cdot\mathrm{P})_{\binom{r}{s}}\begin{pmatrix}\delta r\\ \delta s\end{pmatrix} = \boldsymbol{\tau}^*(\theta)_r(\delta r) + \lambda_x(\mathrm{P})_s(\delta s).$$

Indeed, let us develop the 1-form $\lambda_x(\boldsymbol{\tau}\cdot\mathrm{P})$ defined on $\mathrm{U}\times\mathrm{V}$,

$$\begin{aligned}\lambda_x(\boldsymbol{\tau}\cdot\mathrm{P})_{\binom{r}{s}}\begin{pmatrix}\delta r\\ \delta s\end{pmatrix} &= \lambda_x[(r,s)\mapsto(\boldsymbol{\tau}(r),\mathrm{P}(s))\mapsto\tau(r)\cdot\mathrm{P}(s)]_{\binom{r}{s}}\begin{pmatrix}\delta r\\ \delta s\end{pmatrix}\\ &= [\lambda_x^{\mathrm{U},s}(r)\quad\lambda_x^{\mathrm{V},r}(s)]\begin{pmatrix}\delta r\\ \delta s\end{pmatrix},\end{aligned}$$

because every 1-form on $\mathrm{U}\times\mathrm{V}$ at a point $(r,s)\in\mathrm{U}\times\mathrm{V}$ writes $[\lambda_x^{\mathrm{U},s}(r)\,\lambda_x^{\mathrm{V},r}(s)]$, where $\lambda_x^{\mathrm{U},s}$ is a 1-form on U depending on $s$, and $\lambda_x^{\mathrm{V},r}$ is a 1-form on V depending on $r$. Let us use indifferently $\tau_r=\boldsymbol{\tau}(r)$, and $y_s=\mathrm{P}(s)$. We have

$$\begin{aligned}\lambda_x(\boldsymbol{\tau}\cdot\mathrm{P})_{\binom{r}{s}}\begin{pmatrix}\delta r\\ \delta s\end{pmatrix} &= \lambda_x^{\mathrm{U},s}(r)(\delta r)+\lambda_x^{\mathrm{V},r}(s)(\delta s)\\ &= \lambda_x[r\mapsto\boldsymbol{\tau}(r)\cdot y_s]_r(\delta r)+\lambda_x[s\mapsto\tau_r\cdot\mathrm{P}(s)]_s(\delta s)\\ &= \lambda_x(\hat{y}_s\circ\boldsymbol{\tau})_r(\delta r)+\tau_r^*(\lambda)(\mathrm{P})_s(\delta s)\\ &= \boldsymbol{\tau}^*[\hat{y}_s^*(\lambda)]_r(\delta r)+\lambda(\mathrm{P})_s(\delta s)\\ &= \boldsymbol{\tau}^*(\theta)_r(\delta r)+\lambda(\mathrm{P})_s(\delta s).\end{aligned}$$

Therefore,

$$\Phi_x^*(\lambda_x)(r\mapsto y_r)_r(\delta r)=\boldsymbol{\tau}^*(\theta)_r(\delta r)+\lambda(r\mapsto y_r)_r(\delta r),$$

where $\boldsymbol{\tau}(r)=\tau(y_r)$. Now, having $\Phi_x^*(\lambda_x)=\lambda_x$ means that $\Phi_x^*(\lambda_x)(r\mapsto y_r)_r(\delta r)=\lambda_x(r\mapsto y_r)_r(\delta r)$, for all plots $r\mapsto y_r$. That is, $\boldsymbol{\tau}^*(\theta)_r(\delta r)=0$ for all plots $r\mapsto y_r$. Locally, the plot $r\mapsto\boldsymbol{\tau}(r)=\tau(y_r)$ lifts to $\mathbf{R}$ into a plot $r\mapsto t(y_r)$, such that $\tau(y_r)=\mathrm{class}(t(y_r))$, where $\mathrm{class}:\mathbf{R}\to\mathrm{T}_x=\mathbf{R}/\mathrm{P}_\omega$. Then, $\boldsymbol{\tau}^*(\theta)_r=[r\mapsto t(y_r)]^*(dt)=\mathrm{d}[r\mapsto(t(y_r))]$. And this must be zero. Thus $r\mapsto t(y_r)$ is locally constant, and so is $r\mapsto\tau(y_r)$, for all plots $r\mapsto y_r$. Hence, $y\mapsto\tau(y)$ is locally constant, and since $\mathrm{Y}_x$ is connected, $y\mapsto\tau(y)$ is constant, equal to some $\tau_x$. □

Now, the morphism $\Phi$ writes $\Phi(y)=\tau_x\cdot y$, with $x=\mathrm{src}(y)$. Applying the groupoid morphism rule $\Phi(y\cdot y')=\Phi(y)\cdot\Phi(y')$, we get, with $x=\mathrm{src}(y)$ and $x'=\mathrm{trg}(y)=\mathrm{src}(y')$,

$$\tau_x\cdot y\cdot y'=\tau_x\cdot y\cdot\tau_{x'}\cdot y'\quad\Rightarrow\quad\tau_{x'}=\mathbf{1}_{x'}.$$

Therefore, for all $y$ and $x=\mathrm{src}(y)$, $\Phi(y)=y$. The only automorphism of the structure $(\mathbf{T}_\omega,\boldsymbol{\lambda})$ projecting onto the identity of X is the identity of $\mathscr{Y}$, and thus the projection $\mathrm{pr}_2:\mathrm{Aut}(\mathbf{T}_\omega,\boldsymbol{\lambda})\to\mathrm{Diff}(\mathrm{X},\omega)$ is an isomorphism. □

## V. Remarks and Applications

This section discusses the significance and implications of the main theorem, structured into several articles.

**18. On the System: The Prequantum Groupoid as the Central Object.**

The construction presented in this paper introduces a shift in the geometric quantization program by positioning the prequantum groupoid $(\mathbf{T}_\omega, \boldsymbol{\lambda})$ as the central object, in place of the traditional prequantum principal bundle.

This groupoid object is an intrinsic construction, derived solely from the parasymplectic space $(X, \omega)$, as a diffeological quotient of the space of paths, a feature that incidentally resonates with Feynman's path integral approach to quantization. This single object serves as a unified geometric container possessing several key virtues:

$*$ **It contains the original classical system** $(X, \omega)$: The space of objects $\mathrm{Obj}(\mathbf{T}_\omega)$ is X, and the fundamental relationship defined by the 2-form $\omega$ is encoded in the curvature of the prequantum 1-form $\boldsymbol{\lambda}$ on the space of morphisms $\mathrm{Mor}(\mathbf{T}_\omega)$.

$*$ **It embodies the prequantum total space**: The space of morphisms $\mathscr{Y} = \mathrm{Mor}(\mathbf{T}_\omega)$ carries the fundamental prequantum 1-form $\boldsymbol{\lambda}$, which serves as the potential for $\omega$ according to the formula $d\boldsymbol{\lambda} + \mathrm{ends}^*(\omega \ominus \omega) = 0$.

$*$ **It faithfully represents the full symmetry group Diff(X, ω)**: As shown in Article 17, the group of $\omega$-preserving diffeomorphisms of X acts as the full group of automorphisms of the entire $(\mathbf{T}_\omega, \boldsymbol{\lambda})$ structure in a faithful manner, without involving central extensions at this level.

$*$ **It geometrically captures the quantum phase information**: The periodicity associated with $\omega$ is intrinsically represented by the structure of the isotropy groups $\mathbf{T}_{\omega,x}$ at each point $x \in X$ — the "vertical" structure of the groupoid which is isomorphic to the torus of periods $T_\omega$ — as the quotient of the space of loops.

**19. Structure of the Space of Morphisms $\mathscr{Y}$.**

The space of morphisms $\mathscr{Y} = \mathrm{Mor}(\mathbf{T}_\omega)$ possesses a rich internal structure related to the principal bundle slices $\mathscr{Y}_x = \mathrm{Mor}_{\mathbf{T}_\omega}(x, *)$, where $x \in X$ is a fixed base point. Recall from the main theorem (Theorem II, part 1) that $\hat{1}_x : \mathscr{Y}_x \to X$ is a principal $\mathbf{T}_{\omega,x}$-bundle with structure group $\mathbf{T}_{\omega,x} \simeq T_\omega$.

Now, consider the square of the projection $\hat{1}_x : \mathscr{Y}_x \to X$, that is,

$$\hat{1}_x \times \hat{1}_x : \mathscr{Y}_x \times \mathscr{Y}_x \to X \times X, \text{ with } \hat{1}_x \times \hat{1}_x : (y, y') \mapsto (\hat{1}(y), \hat{1}(y')).$$

Since $\hat{1}_x : \mathscr{Y}_x \to X$ is itself a $\mathbf{T}_{\omega,x}$-principal fiber bundle, this is a $\mathbf{T}^2_{\omega,x}$-principal fiber bunlde, for the action $(\tau, \tau') \cdot (y, y') = (\tau \cdot y, \tau' \cdot y')$. According to the reduction of principal bundle described in [PIZ13, §8.18], the quotient of $\mathscr{Y}_x \times \mathscr{Y}_x$ by the diagonal subgroup $\mathbf{T}_{\omega,x} \simeq \{(\tau, \tau) \mid \tau \in \mathbf{T}_{\omega,x}\}$ is itself a fiber bundle over the same base space $X \times X$, by the projection class : $(y, y') \mapsto \mathscr{Y} \times_{\mathbf{T}_{\omega,x}} \mathscr{Y} = (\mathscr{Y} \times \mathscr{Y})/\mathbf{T}_{\omega,x}$, with fiber ${\mathbf{T}_{\omega,x}}^2/\mathbf{T}_{\omega,x} \simeq \mathbf{T}_{\omega,x}$ (as homogeneous space).

Now, the quotient $\mathscr{Y}_x \times_{\mathbf{T}_{\omega,x}} \mathscr{Y}_x$ is naturally realized by the projection

$$\mathrm{class} : \mathscr{Y}_x \times \mathscr{Y}_x \to \mathscr{Y} \text{ with } \mathrm{class}(y, y') = y^{-1} \cdot y'.$$

Therefore, the projection ends : $\mathscr{Y} \to X \times X$ is itself a fiber bundle with homogeneous fiber $T_\omega$, which is summarized by the diagram:

$$\begin{array}{ccc} \mathscr{Y}_x \times \mathscr{Y}_x & \xrightarrow{\text{class}} & \mathscr{Y} \\ & {}_{\hat{1}_x \times \hat{1}_x} \searrow \quad \swarrow_{\text{ends}} & \\ & X \times X & \end{array}$$

**Note.** As a consequence of the groupoid structure, for any fixed path $\gamma_0 \in \mathrm{Paths}(X, x, x')$, the map $\tau \mapsto \tau \cdot [\gamma_0]_\omega$ establishes a diffeomorphism from the isotropy group $\mathbf{T}_{\omega,x}$ to the fiber $\mathrm{Mor}_{\mathbf{T}_\omega}(x, x')$. This means that $\mathrm{Mor}_{\mathbf{T}_\omega}(x, x')$ is a principal homogeneous space for the isotropy group $\mathbf{T}_{\omega,x} \simeq T_\omega$.

Thus, the fibers of the bundle ends : $\mathscr{Y} \to X \times X$ are all diffeomorphic to the torus of periods $T_\omega$. This holds for any pair of points $(x, x')$.

When the torus of periods $T_\omega = \mathbf{R}/P_\omega$ is a strict diffeological space (i.e., not diffeomorphic to $S^1$ or $\mathbf{R}$), the total space of the groupoid $\mathscr{Y}$ will also be a strict diffeological space, even if the base space X is a smooth manifold. This highlights how the diffeological framework naturally accommodates these structures that arise from the geometry of $\omega$ and may not fit into the traditional manifold setting.

**20. The Exact Case.**

When the closed 2-form $\omega$ is exact, i.e., $\omega = d\alpha$ for some 1-form $\alpha \in \Omega^1(X)$, the unique prequantum groupoid construction simplifies significantly, revealing a particularly direct relationship to the primitive $\alpha$.

In this case, the group of periods $P_\omega$ is trivial, $P_\omega = \{0\}$. Indeed, $\mathsf{K}\omega = \mathsf{K}(d\alpha) = \hat{1}^*(\alpha) - \hat{0}^*(\alpha) - d[\mathsf{K}\alpha]$. Note that $f_\omega := \mathsf{K}\alpha \in \Omega^0(\mathrm{Paths}(X))$ is a smooth map, explicitly:

$$f_\omega : \mathrm{Paths}(X) \to \mathbf{R} \text{ is defined by } f_\omega(\gamma) = \int_\gamma \alpha.$$

Now, restricted to Loops(X), we get $\mathsf{K}\omega \restriction \mathrm{Loops}(X) = df_\omega \restriction \mathrm{Loops}(X)$, and the periods are the integrals $\int_\sigma df_\omega = f_\omega(\sigma(1)) - f_\omega(\sigma(0))$. However, when $\sigma \in \mathrm{Loops}(\mathrm{Loops}(X))$, $f_\omega(\sigma(1)) = f_\omega(\sigma(0))$, which implies $P_\omega = \{0\}$ and $T_\omega = \mathbf{R}/P_\omega = \mathbf{R}$.

Next, the equivalence relation $\gamma \sim_\omega \gamma'$ means: $\mathrm{ends}(\gamma) = \mathrm{ends}(\gamma') = (x, x')$ and $\int_\sigma \mathsf{K}\omega = 0$ for any fixed-ends homotopy $\sigma : s \mapsto \gamma_s$ from $\gamma$ to $\gamma'$. But:

$$\begin{aligned} \int_\sigma \mathsf{K}\omega &= \int_\sigma (\hat{1}^*(\alpha) - \hat{0}^*(\alpha) - d[f_\omega]) = \int_{\hat{1}\circ\sigma} \alpha - \int_{\hat{0}\circ\sigma} \alpha - \int_\sigma df_\omega \\ &= \int_{s\mapsto\gamma_s(1)} \alpha - \int_{s\mapsto\gamma_s(0)} \alpha - (f_\omega(\gamma) - f_\omega(\gamma')) = \underbrace{\int_{s\mapsto x'} \alpha}_{=0} - \underbrace{\int_{s\mapsto x} \alpha}_{=0} - f_\omega(\gamma) + f_\omega(\gamma') \end{aligned}$$

$$= f_\omega(\gamma') - f_\omega(\gamma)$$

Then,

$$\gamma \sim_\omega \gamma' \text{ iff } \operatorname{ends}(\gamma) = \operatorname{ends}(\gamma') \text{ and } \int_\gamma \alpha = \int_{\gamma'} \alpha.$$

The space of morphisms $\mathscr{Y} = \mathrm{Paths}(\mathrm{X})/\sim_\omega$ can be identified diffeomorphically with $\mathrm{X} \times \mathbf{R} \times \mathrm{X}$, by the subduction

$$\mathscr{Y} \simeq \mathrm{X} \times \mathbf{R} \times \mathrm{X} \text{ with } \operatorname{class}_\omega : \gamma \mapsto \left( x = \gamma(0), t = \int_\gamma \alpha, x' = \gamma(1) \right)$$

By additivity of the integral of $\alpha$ on paths, the groupoid composition becomes

$$(x, t, x') \cdot (x', t', x'') = (x, t + t', x")$$

This groupoid is an *additive groupoid* over X. Note that, the identity of the object $x \in \mathrm{X}$ in the groupoid is the arrow $(x, 0, x)$. The map $x \mapsto (x, 0, x)$ from X to $\mathscr{Y}$ is the identity induction.

The prequantum 1-form $\lambda$ on $\mathscr{Y}$, uniquely determined by $\operatorname{class}_\omega^*(\lambda) = \mathsf{K}\omega$, can be expressed on a 1-plot $s \mapsto \gamma_s$ as: $\operatorname{class}_\omega^*(\lambda)(s \mapsto \gamma_s) = \lambda(s \mapsto \operatorname{class}_\omega(\gamma_s)) = \mathsf{K}\omega(s \mapsto \gamma_s)$. That is, $\lambda(s \mapsto (x_s, t_s, x'_s)) = \mathsf{K}\omega(s \mapsto \gamma_s)$, with $x_s = \gamma_s(0)$, $t_s = f_\omega(\gamma_s)$ and $x'_s = \gamma_s(1)$. But,

$$\begin{aligned}
\mathsf{K}\omega(s \mapsto \gamma_s) &= \hat{1}^*(\alpha)(s \mapsto \gamma_s) - \hat{0}^*(\alpha)(s \mapsto \gamma_s) - [d f_\omega](s \mapsto \gamma_s), \\
&= \alpha(s \mapsto \gamma_s(1))) - \alpha(s \mapsto \gamma_s(0))) - d[s \mapsto f_\omega(\gamma_s)], \\
&= \alpha(s \mapsto x'_s) - \alpha(s \mapsto x_s) - d[s \mapsto t_s]
\end{aligned}$$

Therefore,

$$\lambda(s \mapsto (x_s, t_s, x'_s)) = \alpha(s \mapsto x'_s) - \alpha(s \mapsto x_s) - d[s \mapsto t_s]$$

That is, with reordering terms:

$$\lambda = \mathrm{pr}_3^*(\alpha) - \mathrm{pr}_2^*(dt) - \mathrm{pr}_1^*(\alpha),$$

where the $\mathrm{pr}_i$, $i = 1, 2, 3$, are the projections of $\mathscr{Y} \simeq \mathrm{X} \times \mathbf{R} \times \mathrm{X}$ on its three factors.

Since we are in a very simple case, we can easily check that the main properties for $(\mathbf{T}_\omega, \lambda)$ are satisfied:

$*$ **Invariance of $\lambda$ under Groupoid Composition**: We verify the left and right invariance of $\lambda$ using its coordinate expression: $y = (x, t, x') \in \mathscr{Y} \simeq \mathrm{X} \times \mathbf{R} \times \mathrm{X}$.

Let us check the left invariance. Let $y_0 = (x_0, t_0, x'_0) \in \mathscr{Y}$ and $y = (x, t, x')$. Then, $\mathsf{L}(y_0)(y) = (x_0, t_0, x'_0) \cdot (x, t, x')$, which composes only when $x = x'_0$. Then $\mathsf{L}(y_0)(y) = (x_0, t_0, x'_0) \cdot (x'_0, t, x') = (x_0, t_0 + t, x')$, that is, $\mathsf{L}(y_0)(x'_0, t, x') = (x_0, t_0 + t, x')$. Now $\lambda \upharpoonright \mathscr{Y}_{x'_0,*} = \mathrm{pr}_3^*(\alpha) - \mathrm{pr}_2^*(dt)$. Thus $[(x'_0, t, x') \mapsto (x_0, t_0 + t, x']^*((\mathrm{pr}_3^*(\alpha) - \mathrm{pr}_2^*(dt))) = \mathrm{pr}_3^*(\alpha) - \mathrm{pr}_2^*(dt) = \lambda \upharpoonright \mathscr{Y}_{x_0,*}$. The same applies to right invariance.

$*$ **Curvature of $\boldsymbol{\lambda}$**: The differential of $\boldsymbol{\lambda}$ gives $d[\mathrm{pr}_3^*(\alpha)] - d[\mathrm{pr}_2^*(dt)] - d[\mathrm{pr}_1^*(\alpha)] = \mathrm{pr}_3^*(d\alpha) - 0 - \mathrm{pr}_1^*(d\alpha)$, that is, $d\boldsymbol{\lambda} = \mathrm{pr}_3^*(\omega) - \mathrm{pr}_1^*(\omega)$, which was announced by the theory.

$*$ **Action of Diff(X, ω)**: Let $\phi \in \mathrm{Diff}(\mathrm{X}, \omega)$, then $\phi^*(d\alpha) = d\alpha$ implies $d[\phi^*(\alpha)] = d\alpha$, and then $d[\phi^*(\alpha) - \alpha] = 0$. That is $\phi^*(\alpha) - \alpha$ is closed, but since X is simply connected $\phi^*(\alpha) - \alpha$ is exact and there exists a function $f_\phi \in \mathrm{C}^\infty(\mathrm{X}, \mathbf{R})$ such that $\phi^*(\alpha) = \alpha + df_\phi$. So, let

$$\Phi[\gamma]_\omega = [\phi \circ \gamma]_\omega = \left(\phi(\gamma(0)), \int_{\phi\circ\gamma} \alpha, \phi(\gamma(1))\right).$$

But $\int_{\phi\circ\gamma} \alpha = \int_\gamma \phi^*(\alpha) = \int_\gamma \alpha + df_\phi = \int_\gamma \alpha + \int_\gamma f_\phi = \int_\gamma \alpha + f_\phi(\gamma(1)) - f_\phi(\gamma(0))$. Therefore,

$$\Phi(x, t, x') = (\phi(x), t + f_\phi(x') - f_\phi(1), \phi(x')).$$

Hence,

$$\begin{aligned}
\Phi^*(\boldsymbol{\lambda}) &= \Phi^*(\mathrm{pr}_3^*(\alpha) - \mathrm{pr}_2^*(dt) - \mathrm{pr}_1^*(\alpha) \\
&= \Phi^*(\mathrm{pr}_3^*(\alpha)) - \Phi^*(\mathrm{pr}_2^*(dt)) - \Phi^*(\mathrm{pr}_1^*(\alpha)) \\
&= (\mathrm{pr}_3 \circ \Phi)^*(\alpha) - (\mathrm{pr}_2 \circ \Phi)^*(dt) - (\mathrm{pr}_1 \circ \Phi)^*(\alpha) \\
&= (\phi \circ \mathrm{pr}_3)^*(\alpha) - [t + f_\phi(x') - f_\phi(x)]^*(dt) - (\phi \circ \mathrm{pr}_1)^*(\alpha) \\
&= \mathrm{pr}_3^*(\phi^*(\alpha)) - d[t + f_\phi(x') - f_\phi(x)] - \mathrm{pr}_1^*(\phi^*(\alpha)) \\
&= \mathrm{pr}_3^*(\alpha + df_\phi) - dt - d[f_\phi \circ \mathrm{pr}_3] + d[f_\phi \circ \mathrm{pr}_1] - \mathrm{pr}_1^*(\alpha + df_\phi) \\
&= \mathrm{pr}_3^*(\alpha) - dt - \mathrm{pr}_1^*(\alpha) + \mathrm{pr}_3^*(df_\phi) - d[\mathrm{pr}_3^*(f_\phi)] + d[\mathrm{pr}_1^*(f_\phi)] - \mathrm{pr}_1^*(df_\phi) \\
&= \mathrm{pr}_3^*(\alpha) - dt - \mathrm{pr}_1^*(\alpha) + \mathrm{pr}_3^*(df_\phi) - \mathrm{pr}_3^*(df_\phi) + \mathrm{pr}_1^*(df_\phi) - \mathrm{pr}_1^*(df_\phi) \\
&= \mathrm{pr}_3^*(\alpha) - dt - \mathrm{pr}_1^*(\alpha) \\
&= \boldsymbol{\lambda}.
\end{aligned}$$

This tedious computation, confirming that Diff(X, ω) acts on $\mathscr{Y}$ while preserving $\boldsymbol{\lambda}$, is not without purpose. It shows how the function cocycle $f_\phi$ of Diff(X, ω), related to the lack of invariance of the primitive α, is compensated by the variance of the isotropy such that Diff(X, ω) acts on $\mathbf{T}_\omega$ as an automorphism of $(\mathbf{T}_\omega, \boldsymbol{\lambda})$.

**Note.** In this exact case, the prequantum bundles $\mathscr{Y}_x$ are trivial **R**-bundles, mirroring the classical situation for exact symplectic forms on manifolds. This recovery of the classical structure in the simplest case demonstrates the consistency and generality of the path-space construction.

The connection form $\lambda = \boldsymbol{\lambda} \restriction \mathscr{Y}_x$ corresponds to a connection on the trivial **R**-bundle $\mathrm{X} \times \mathbf{R} \to \mathrm{X}$ whose curvature is ω.

**21. Example: The Prequantum Groupoid for** $(\mathrm{S}^2, \omega_{S^2})$**.**

The 2-sphere $S^2$, equipped with its standard symplectic form $\omega_{S^2}$, is a fundamental example in geometric quantization. It is a connected and simply connected manifold, and $\omega_{S^2}$ is a closed 2-form. The periods of $\omega_{S^2}$ are related to $\pi_2(S^2) \simeq \mathbf{Z}$. For a suitable normalization (e.g., such that $\int_{S^2} \omega_{S^2} = 2\pi$), the group of periods $P_{\omega_{S^2}}$ is $2\pi\mathbf{Z}$. This is a discrete subgroup of $\mathbf{R}$. Thus, $(S^2, \omega_{S^2})$ satisfies the conditions of the main theorem.

The torus of periods is $T_{\omega_{S^2}} = \mathbf{R}/P_{\omega_{S^2}} = \mathbf{R}/2\pi\mathbf{Z} \simeq S^1$.

According to the main theorem, the prequantum groupoid $\mathbf{T}_{\omega_{S^2}}$ for $(S^2, \omega_{S^2})$ has $S^2$ as its objects. The space of morphisms is $\mathscr{Y} = \mathrm{Paths}(S^2)/\sim_{\omega_{S^2}}$, where the equivalence relation $\sim_{\omega_{S^2}}$ is determined by $\omega_{S^2}$ and its periods $2\pi\mathbf{Z}$. The isotropy group at any point $x \in S^2$ is isomorphic to the torus of periods: $\mathbf{T}_{\omega_{S^2},x} \simeq T_{\omega_{S^2}} \simeq S^1$.

The space of morphisms $\mathscr{Y}$ has a rich structure over the product space $S^2 \times S^2$ via the source/target map ends : $\mathscr{Y} \to S^2 \times S^2$. As discussed in Article 19, this map is a diffeological fiber bundle with fiber $T_{\omega_{S^2}} \simeq S^1$.

In this specific case, the prequantum line bundle over $S^2$ (corresponding to the prequantization condition $[\omega_{S^2}] \in \mathsf{H}^2_{\mathrm{dR}}(S^2)$ being integral) is the well-known Hopf bundle $\pi : S^3 \to S^2$, which is a principal $S^1$-bundle. The space of morphisms $\mathscr{Y}$ of the prequantum groupoid $\mathbf{T}_{\omega_{S^2}}$ can be identified with the space of isomorphisms between fibers of this principal bundle. This space is diffeomorphic to $(S^3 \times S^3)/S^1_{\mathrm{diag}}$, where $S^1_{\mathrm{diag}}$ is the diagonally embedded subgroup in $S^1 \times S^1$.

The map ends : $\mathscr{Y} \to S^2 \times S^2$ is a $S^1$-bundle over $S^2 \times S^2$. The base space $S^2 \times S^2$ is a 4-dimensional manifold, the fiber is $S^1$ (a 1-dimensional manifold), and the total space $\mathscr{Y} \simeq (S^3 \times S^3)/S^1_{\mathrm{diag}}$ is a 5-dimensional manifold. The prequantum 1-form $\boldsymbol{\lambda}$ on $\mathscr{Y}$ is a 1-form on this 5-manifold whose curvature $d\boldsymbol{\lambda}$ is related to $\omega_{S^2} \ominus \omega_{S^2}$ on the base $S^2 \times S^2$.

This example provides a concrete illustration of the prequantum groupoid construction for a classical symplectic manifold, showing how the space of morphisms $\mathscr{Y}$ forms a principal $S^1$-homogeneous-bundle over the product of the base space with itself, with the expected dimensions and structure related to the classical prequantum bundle.

**Note.** For any pair of non-antipodal points $(x, x') \in S^2 \times S^2$, the space of morphisms $\mathrm{Mor}_{\mathbf{T}_{\omega_{S^2}}}(x, x')$, diffeomorphic to $S^1$, can be trivialized. The trivialization is explicitly realized by mapping a class $[\gamma]_{\omega_{S^2}}$ (where $\gamma$ is a path from $x$ to $x'$) as follows

$$[\gamma]_{\omega_{S^2}} = \left(x, \mathrm{e}^{i\int_{\gamma_{x,x'}}^{\gamma} \mathsf{K}\omega_{S^2}}, x'\right),$$

where $\gamma_{x,x'}$ is the unique shortest geodesic from $x$ to $x'$, and the integral is taken over any fixed-ends homotopy between $\gamma_{x,x'}$ and $\gamma$ in Paths($S^2$, $x$, $x'$). This corresponds to the integral of $\omega_{S^2}$ on any surface sweeped by the homotopy. This identification is a local trivialization of $\mathscr{Y}_{S^2}$ over the subset of non-antipodal points of the sphere. It connects with the physical intuition often found in the literature, between the abstract path-space quotient and a concrete phase value associated with paths between points.

**22. Application to Symplectic Reduction.**

A central and often problematic question in geometric quantization is whether the process of quantization commutes with symplectic reduction.[11] Classically, quantizing a symplectic manifold and then performing a reduction (Quantize then Reduce, QTR) frequently does not yield the same result as first reducing the classical system and then quantizing the reduced space (Reduce then Quantize, RTQ). This discrepancy is particularly acute when the reduced space is singular, lacking a smooth manifold structure. Traditional geometric quantization methods, heavily reliant on local Euclidean charts, face significant obstacles when applied directly to such singular reduced spaces.

The diffeological framework, by providing a robust setting for differential geometry on arbitrary spaces including those with singularities, allows us to construct the prequantum groupoid $(\mathbf{T}_\omega, \boldsymbol{\lambda})$ even for singular parasymplectic spaces arising from reduction. This holistic approach, which places the prequantum groupoid as a unified object encoding both classical and prequantum information along with the full symmetry group, offers a novel perspective on the reduction problem. This perspective resonates with the general philosophy of diffeology regarding quotient spaces, as discussed in [GIZ25].

When considering a quotient procedure $X/\sim$, diffeology encourages focusing on the resulting quotient space itself, equipped with the quotient diffeology, rather than being tied to the specific procedure that generated it. For instance, the Hopf reduction $S^3/S^1$ is naturally viewed as the smooth 2-sphere $S^2$ with its standard manifold structure (which is a particular diffeology). While the Hopf fibration can be retrieved by considering the set of $S^1$ principal bundles over $S^2$ (classified by Chern classes), the primary object of interest is the resulting space $S^2$.

The challenge arises when the quotient space possesses singularities and cannot be naturally equipped with a compatible manifold structure. In such cases, the resulting space might, at best, be considered a topological space, as exemplified by the quotient $\mathcal{Q}_m = \mathbf{C}/\mathbf{Z}_m$. Topologically, this space is homeomorphic to $\mathbf{C}$. However, this topological perspective fails to capture the richer differential structure that is intuitively present, particularly at the singularity.

[11]Reflecting on the history of this subject, I recall giving my first talk at Souriau's seminar more than 40 years ago, specifically on Alan Weinstein's seminal paper on symplectic reduction. I still vividly remember it was a Tuesday afternoon.

This is precisely the motivation behind notions like V-manifolds (Thurston's orbifolds), introduced by Ichiro Satake [Sat56] and shown to be equivalent to the specific class of diffeological orbifolds [IKZ10]. Diffeology provides a framework where, when constructing a quotient, one can indeed focus on the resulting quotient diffeological space, regardless of the specific process. This applies to quotients of manifolds by compact groups, as shown in [GIZ25], and is relevant to problems of symplectic reduction, generally denoted $\mathrm{J}^{-1}(0)/\mathrm{G}$.[12]

**Diffeology Principle:** A quotient $\mathcal{Q} = \mathrm{X}/\sim$ is always regarded as a quotient diffeological space, irrespective of its construction.

Applying this principle to the symplectic reduction $\mathbf{C}/\mathbf{Z}_m$, the resulting space is the cone-orbifold $\mathcal{Q}_m$. It has been proved that $\mathcal{Q}_m$ is *symplectically generated* in the diffeological sense;[13] that is, there exists a closed 2-form $\omega$ (a parasymplectic form) on $\mathcal{Q}_m$ such that its pullback on $\mathbf{C}$ by the projection $\pi_m : z \mapsto [z]$ is the standard symplectic form $dx \wedge dy$ [PIZ13, §9.32].[14] Actually, every parasymplectic form on $\mathcal{Q}_m$ is proportional to this $\omega$ by a smooth function on $\mathcal{Q}_m$.

This parasymplectic diffeological space $(\mathcal{Q}_m, \omega)$ is connected and simply connected; it is, in fact, contractible. The map $\rho_s : [z] \mapsto [sz]$, where $s \in \mathbf{R}$ and $[z] \in \mathcal{Q}_m$, is well-defined and provides a deformation retraction from $\mathcal{Q}_m$ (for $s = 1$) to the origin $\{[0]\}$ (for $s = 0$). Since $\mathcal{Q}_m$ is contractible, the parasymplectic form $\omega$ is exact [PIZ13, §6.90]. Let $\alpha \in \Omega^1(\mathcal{Q}_m)$ be a primitive such that $\omega = d\alpha$. We are thus in the case previously discussed in Article 20. The result applies directly:

**Proposition.** *The prequantum groupoid $\boldsymbol{\mathcal{Q}_{m,\omega}}$ is isomorphic to the additive groupoid*

$$\boldsymbol{\mathcal{Q}_{m,\omega}} \simeq \{(q,t,q') \in \mathcal{Q}_m \times \mathbf{R} \times \mathcal{Q}_m\} \;\; \textit{with} \;\; (q,t,q') \cdot (q',t',q'') = (q, t+t', q'').$$

*The prequantum form* $\boldsymbol{\lambda}$ *is given by* $\boldsymbol{\lambda} = \mathrm{pr}_3^*(\alpha) - \mathrm{pr}_1^*(\alpha) - \mathrm{pr}_2^*(dt)$.

[12] The map J is the moment map of a Hamiltonian equivariant action of a Lie group G. The group G acts on the level set $\mathrm{J}^{-1}(0)$, where the restriction of the symplectic form $\omega$ is coisotropic. If $\mathrm{J}^{-1}(0)$ is a manifold such that the canonical projection to the quotient is a submersion, then the form $\omega \restriction \mathrm{J}^{-1}(0)$ descends to the quotient as a symplectic form. This is the Marsden-Weinstein construction [MW74].

[13] A parasymplectic space $(\mathrm{X}, \omega)$ is symplectically generated if there is a generating family [PIZ13, §1.66] of plots where the pullback of $\omega$ is symplectic.

[14] One can choose a projection $\pi_m(z) = z^m$, in which case $\mathcal{Q}_m$ is diffeomorphic to $\mathbf{C}$ equipped with the pushforward of the standard smooth diffeology on $\mathbf{C}$ by this map $\pi_m$.

This completely addresses the prequantization of the parasymplectic cone-orbifold $(\mathcal{Q}_m, \omega)$ within the diffeological framework, thereby obviating the need to directly address the commutation of quantization and reduction procedures at the level of the base space.

**Note.** A critical insight from this explicit construction is that the singular origin $0 \in \mathcal{Q}_m$ does not lead to a degeneration or alteration of the isotropy group. For this exact case, the isotropy group $\mathbf{T}_{\omega,q}$ is isomorphic to $\mathbf{R}$ for all points $q \in \mathcal{Q}_m$. This uniformity of the quantum phase information arises naturally from the inherent structure of the path-space quotient, as discussed in the Introduction (Note 2 of Article 8), demonstrating that singularities in the base space do not intrinsically affect the nature of the isotropy group itself. If quantum phenomena are related to such singularities, it would rather be through the representations of the group of automorphisms (which reveal the singular structure via the Klein stratification) than through the local isotropy itself.

**23. Example: The Prequantum Groupoid for Loops($S^3$).**

The construction of the prequantum groupoid $(\mathbf{T}_\omega, \boldsymbol{\lambda})$ is applicable not only to singular spaces but also to infinite-dimensional diffeological spaces, provided they satisfy the conditions of the main theorem (connected, simply connected, with a closed 2-form having discrete periods). A relevant example is the loop space of the three-sphere, $X = \mathrm{Loops}(S^3)$.

The space $X = \mathrm{Loops}(S^3)$ is an infinite-dimensional diffeological space. Its path-connectedness is given by $\pi_0(\mathrm{Loops}(S^3)) \simeq \pi_1(S^3) = \{0\}$; thus X is connected. Its simple connectedness is given by $\pi_1(\mathrm{Loops}(S^3)) \simeq \pi_2(S^3) = \{0\}$; thus X is simply connected.

Consider the volume 3-form vol on $S^3$. Let $\omega = \mathsf{K}\,\mathrm{vol} \in \Omega^2(X)$ be the 2-form on X obtained by applying the chain-homotopy operator. The 2-form $\omega$ is closed, which follows from $d\,\mathrm{vol} = 0$ and the properties of the chain-homotopy operator. The group of periods $P_\omega$ of the form $\omega$ is the group of periods of the 1-form $\mathsf{K}\omega \upharpoonright \mathrm{Loops}(\mathrm{Loops}(X))$. This group is a homomorphic image of $\pi_1(\mathrm{Loops}(X)) \simeq \pi_2(X) \simeq \pi_3(S^3) = \mathbf{Z}$. The periods of $\omega$ are related to integrals of vol over cycles in $\pi_3(S^3)$. For a suitable normalization of the volume form vol on $S^3$ (e.g., such that $\int_{S^3} \mathrm{vol} = 1$), the group of periods $P_\omega$ is isomorphic to $\mathbf{Z}$.

The torus of periods is $T_\omega = \mathbf{R}/P_\omega = \mathbf{R}/\mathbf{Z} \simeq S^1$.

According to the main theorem, the prequantum groupoid $\mathbf{T}_\omega$ for $(X, \omega)$ has $X = \mathrm{Loops}(S^3)$ as its objects. The space of morphisms is $\mathscr{Y} = \mathrm{Paths}(X)/\sim_\omega = \mathrm{Paths}(\mathrm{Loops}(S^3))/\sim_\omega$, where the equivalence relation $\sim_\omega$ is determined by the 2-form $\omega$ and its periods $P_\omega = \mathbf{Z}$.

The isotropy group at any object $\ell \in X$ (i.e., at any loop $\ell \in \mathrm{Loops}(S^3)$) is isomorphic to the torus of periods: $\mathbf{T}_{\omega,\ell} \simeq T_\omega \simeq S^1$. This means that at each point in the infinite-dimensional space $\mathrm{Loops}(S^3)$, the quantum phase information is captured by an $S^1$ group.

The prequantum 1-form $\boldsymbol{\lambda}$ on $\mathscr{Y}$ is the unique form such that $\mathrm{class}_\omega^*(\boldsymbol{\lambda}) = \mathsf{K}\omega$.

Explicitly representing the space of morphisms $\mathscr{Y} = \mathrm{Paths}(\mathrm{Loops}(S^3))/\sim_\omega$ in a simpler form (like a product space) is generally not possible due to the infinite-dimensional nature of X and the non-exactness of $\omega$ (since $P_\omega = \mathbf{Z} \neq \{0\}$). The space $\mathscr{Y}$ is genuinely the quotient space defined by the equivalence relation.

This example illustrates how the prequantum groupoid construction extends to infinite-dimensional settings. The situation shares some analogy with the prequantization of the 2-sphere $(S^2, \omega_{S^2})$, where $S^2$ is connected and simply connected, and the periods of the standard symplectic form are related to $\pi_2(S^2) \simeq \mathbf{Z}$, leading to an $S^1$ isotropy group. The key difference here is the infinite dimensionality of the base space $X = \mathrm{Loops}(S^3)$.

**24. The Moment Map on the Prequantum Groupoid.**

Now that we have this symmetry group $\mathrm{Aut}(\mathbf{T}_\omega, \boldsymbol{\lambda})$, we can define a moment map using its action, and here are its properties and what it tells us.

The prequantum groupoid $(\mathbf{T}_\omega, \boldsymbol{\lambda})$ provides a geometric framework to study the system $(X, \omega)$. As established in Article 16, its group of automorphisms $G_\omega = \mathrm{Aut}(\mathbf{T}_\omega, \boldsymbol{\lambda})$ is diffeologically isomorphic to the group of $\omega$-preserving diffeomorphisms of X, $\mathrm{Diff}(X, \omega)$. This group $G_\omega$ acts smoothly on the space of morphisms $\mathscr{Y} = \mathrm{Mor}(\mathbf{T}_\omega)$, preserving the prequantum 1-form $\boldsymbol{\lambda}$.

According to the general theory of moment maps in diffeology developed in [PIZ10] (see also [PIZ13, Chap. 9]), the existence of a $G_\omega$-invariant 1-form $\boldsymbol{\lambda}$ on $\mathscr{Y}$ ensures the existence of a moment map $\Psi_\omega : \mathscr{Y} \to \mathscr{G}_\omega^*$, where $\mathscr{G}_\omega^*$ is the space of left-invariant 1-forms on the diffeological group $G_\omega$ (the space of momenta [PIZ10, Art. 2.6]). This map is defined for $y \in \mathscr{Y}$ by $\Psi_\omega(y) = \hat{y}^*(\boldsymbol{\lambda})$, where $\hat{y} : G_\omega \to \mathscr{Y}$ is the orbit map $\hat{y}(\Phi) = \Phi(y)$.

This moment map on the space of morphisms $\mathscr{Y}$ is directly related to the paths moment map defined in [PIZ10, Art. 3.1] for the action of $G_\omega$ on X. Let $\Psi_{\mathrm{paths}} : \mathrm{Paths}(X) \to \mathscr{G}_\omega^*$ be the paths moment map given by $\Psi_{\mathrm{paths}}(\gamma) = \hat{\gamma}^*(\mathsf{K}\omega)$, where $\hat{\gamma}(\phi) = \phi \circ \gamma$. The map $\Psi_\omega$ is precisely the projection of $\Psi_{\mathrm{paths}}$ onto the quotient $\mathscr{Y}$:

**Proposition.** *The moment map* $\Psi_\omega : \mathscr{Y} \to \mathscr{G}_\omega^*$ *satisfies* $\Psi_\omega \circ \mathrm{class}_\omega = \Psi_{paths}$.

*Proof.* Let $\gamma \in \mathrm{Paths}(X)$, and let $y = \mathrm{class}_\omega(\gamma) \in \mathscr{Y}$. By definition of $\Psi_\omega$, $\Psi_\omega(y) = \hat{y}^*(\boldsymbol{\lambda})$, where $\hat{y} : G_\omega \to \mathscr{Y}$ is the orbit map $\hat{y}(\Phi) = \Phi(y)$ for $\Phi \in G_\omega$. Recalling that $G_\omega = \mathrm{Aut}(\mathbf{T}_\omega, \boldsymbol{\lambda})$ is isomorphic to $\mathrm{Diff}(X, \omega)$ via $\Phi \mapsto \phi$ such that $\Phi([\gamma']_\omega) = [\phi \circ \gamma']_\omega$, the orbit map $\hat{y}$ is given by $\hat{y}(\Phi) = [\phi \circ \gamma]_\omega = \mathrm{class}_\omega(\phi \circ \gamma)$. Thus, $\hat{y} = \mathrm{class}_\omega \circ \hat{\gamma}$, where $\hat{\gamma} : G_\omega \to \mathrm{Paths}(X)$ is the orbit map $\hat{\gamma}(\phi) = \phi \circ \gamma$ as defined for $\Psi_{\mathrm{paths}}$. Now

we compute the pullback:

$$\Psi_\omega(\mathrm{class}_\omega(\gamma)) = \hat{y}^*(\boldsymbol{\lambda}) = (\mathrm{class}_\omega \circ \hat{\gamma})^*(\boldsymbol{\lambda}) = \hat{\gamma}^*(\mathrm{class}_\omega^*(\boldsymbol{\lambda})).$$

By definition of $\boldsymbol{\lambda}$ (Theorem II, part 3), $\mathrm{class}_\omega^*(\boldsymbol{\lambda}) = \mathsf{K}\omega$. Substituting this gives:

$$\Psi_\omega(\mathrm{class}_\omega(\gamma)) = \hat{\gamma}^*(\mathsf{K}\omega).$$

The right side is precisely $\Psi_{\mathrm{paths}}(\gamma)$. Therefore, $\Psi_\omega \circ \mathrm{class}_\omega = \Psi_{\mathrm{paths}}$. □

The paths moment map $\Psi_{\mathrm{paths}}$ is an additive homomorphism from the path concatenation to $(\mathscr{G}_\omega^*, +)$ [PIZ10, Art. 4.3]. Since $\Psi_\omega$ factors through $\mathrm{class}_\omega$ and $\mathrm{class}_\omega$ preserves concatenation, $\Psi_\omega$ is an additive homomorphism from the groupoid $\mathbf{T}_\omega$ to $(\mathscr{G}_\omega^*, +)$:

$$\Psi_\omega(y \cdot y') = \Psi_\omega(y) + \Psi_\omega(y')$$

for any composable $y, y' \in \mathscr{Y}$.

The additive property $\Psi_\omega(y \cdot y') = \Psi_\omega(y) + \Psi_\omega(y')$ implies that the value of $\Psi_\omega(y)$ for $y \in \mathscr{Y}$ depends only on the endpoints of $y$. This means $\Psi_\omega$ descends to a map $\psi_\omega : \mathrm{X} \times \mathrm{X} \to \mathscr{G}_\omega^*$. This descent occurs if and only if $\Psi_\omega$ is constant on the equivalence classes $[\gamma]_\omega$ corresponding to loops, which is equivalent to $\Psi_\omega(\ell) = 0$ for all loops $\ell \in \mathrm{Loops}(\mathrm{X})$ (viewed as elements of $\mathscr{Y}$ via $\mathrm{class}_\omega$). The set of values $\Psi_{\mathrm{paths}}(\ell) = \Psi_\omega(\mathrm{class}_\omega(\ell))$ for $\ell \in \mathrm{Loops}(\mathrm{X})$ is the universal holonomy group $\Gamma_\omega = \Psi_{\mathrm{paths}}(\mathrm{Loops}(\mathrm{X}))$ for the action of $\mathrm{G}_\omega$ on X [PIZ10, Art. 3.7].

Since X is connected and simply connected, its fundamental group $\pi_1(\mathrm{X})$ is trivial. The universal holonomy group $\Gamma_\omega$ is an additive subgroup of $\mathscr{G}_\omega^*$ which is a homomorphic image of $\pi_0(\mathrm{Loops}(\mathrm{X}))$ [PIZ10, Art. 3.7]. Since X is simply connected and connected, $\pi_0(\mathrm{Loops}(\mathrm{X}))$ is trivial. Therefore, for a simply connected X, the universal holonomy $\Gamma_\omega$ is trivial, $\Gamma_\omega = \{0\}$. This property will not be preserved in the general case of $\pi_1(\mathrm{X}) \neq \{0\}$. In the general case, the universal holonomy $\Gamma_\omega$ will typically be non-trivial.

The fact that $\Gamma_\omega = \{0\}$ for simply connected X implies that $\Psi_\omega(\ell) = 0$ for all loops $\ell \in \mathrm{Loops}(\mathrm{X})$, and thus $\Psi_\omega : \mathscr{Y} \to \mathscr{G}_\omega^*$ descends to a well-defined map on $\mathrm{X} \times \mathrm{X}$. This descended map is the *two-points moment map*:

$$\psi_\omega : \mathrm{X} \times \mathrm{X} \to \mathscr{G}_\omega^* \quad \text{defined by} \quad \psi_\omega(x, x') = \Psi_\omega(y),$$

for any $y \in \mathscr{Y}$ with $\mathrm{ends}(y) = (x, x')$. This $\psi_\omega$ is the universal two-points moment map associated with $\omega$ (compare [PIZ10, Art. 4.1]). It satisfies the Chasles cocycle relation $\psi_\omega(x, x'') = \psi_\omega(x, x') + \psi_\omega(x', x'')$ for $x, x', x'' \in \mathrm{X}$.[15]

From the two-points moment map $\psi_\omega$, we can derive the *one-point moment map* $\mu_\omega : \mathrm{X} \to \mathscr{G}_\omega^*$ by choosing a base point $x_0 \in \mathrm{X}$. As shown in [PIZ10, Art. 5.1],

[15] This moment map $\psi_\omega$ can be interpreted as the moment map for the diffeological space $\mathrm{X} \times \mathrm{X}$ equipped with the 2-form $\omega \ominus \omega = \mathrm{pr}_1^* \omega - \mathrm{pr}_2^* \omega$ and the diagonal action of $\mathrm{G}_\omega$.

$\mu_\omega$ is a primitive of $\psi_\omega$, unique up to an additive constant $\varepsilon \in \mathscr{G}^*_\omega$, satisfying $\psi_\omega(x, x') = \mu_\omega(x') - \mu_\omega(x)$. A standard choice is $\mu_\omega(x) = \psi_\omega(x_0, x)$.

The equivariance properties of these moment maps and the associated Souriau's cocycle $\theta_\omega \in C^\infty(G_\omega, \mathscr{G}^*_\omega)$ are described in detail in [PIZ10, Art. 5.2]. For simply connected X, the holonomy $\Gamma_\omega$ is trivial, which means the target space for Souriau's cocycle and moment maps is $\mathscr{G}^*_\omega$ (as opposed to $\mathscr{G}^*_\omega/\Gamma_\omega$ in the general case). The universal Souriau class $\sigma_\omega$, which lives in $H^1(G_\omega, \mathscr{G}^*_\omega/\Gamma_\omega)$, is therefore an element of $H^1(G_\omega, \mathscr{G}^*_\omega)$. The triviality of this cohomology class implies the existence of an equivariant moment map $\mu_\omega$ *op. cit.* However, the triviality of $\Gamma_\omega$ does not in general imply the triviality of the entire cohomology group $H^1(G_\omega, \mathscr{G}^*_\omega)$. Thus, for a simply connected X, while the holonomy is trivial, Souriau's class $\sigma_\omega$ may be non-trivial, corresponding to non-trivial central extensions in the classical picture.

**25. Towards the Quantum Hilbert Space.**

A central aim of geometric quantization is the construction of the quantum Hilbert space $\mathscr{H}$. In our groupoid framework, the space of morphisms $\mathscr{Y} = \mathrm{Mor}(\mathbf{T}_\omega)$ of the prequantum groupoid $\mathbf{T}_\omega$ serves as the fundamental arena for potential quantum states.

Inspired by the algebraic structure of the groupoid, a natural candidate for quantum states arises from considering *multiplicative functions* on the space of morphisms. A V-valued function $\psi : \mathscr{Y} \to V$, where V is a complex vector space equipped with a suitable composition operation (typically representing the fiber of an associated vector bundle), is called multiplicative if for any composable pair $y_1, y_2 \in \mathscr{Y}$, we have:

$$\psi(y_1 \cdot y_2) = \psi(y_1) \cdot \psi(y_2).$$

Here, the dot $\cdot$ on the left is the groupoid composition, and the dot $\cdot$ on the right is the composition in V. The set of such functions, satisfying appropriate differentiability or regularity conditions, forms a candidate space for the quantum Hilbert space $\mathscr{H}$.

This definition directly encodes the action and representation of the isotropy group $T_\omega$. For any $y \in \mathscr{Y}$ and any $\tau \in \mathbf{T}_{\omega,\mathrm{src}(y)}$ (viewed as an element of $\mathscr{Y}$ with $\mathrm{src}(\tau) = \mathrm{trg}(\tau) = \mathrm{src}(y)$), the composition $\tau \cdot y$ is defined. The multiplicative property then implies $\psi(\tau \cdot y) = \psi(\tau)\psi(y)$. The restriction of $\psi$ to each isotropy group $\mathbf{T}_{\omega,x} \simeq T_\omega$ defines a representation $\rho_x : T_\omega \to \mathrm{End}(V)$ given by $\rho_x(\tau)v = \psi(\tau)v$. For the standard case where $V = \mathbf{C}$, this means $\psi \restriction \mathbf{T}_{\omega,x}$ must be a character (a 1-dimensional representation) of $T_\omega$ for each $x$. Thus, the space of multiplicative functions on the groupoid naturally corresponds to representations of the isotropy group.

As is standard in groupoid theory, the space of multiplicative functions $\psi : \mathscr{Y} \to \mathrm{V}$ satisfying $\psi(\tau \cdot y) = \rho_x(\tau)\psi(y)$ for $\tau \in \mathbf{T}_{\omega,\mathrm{src}(y)}$ is precisely how one identifies sections of the associated vector bundle $\mathscr{E} = \mathscr{Y} \times_{\mathrm{T}_\omega} \mathrm{V} \to \mathrm{X}$ (where the bundle projection is induced by the source map $\mathrm{src} : \mathscr{Y} \to \mathrm{X}$) with equivariant functions on $\mathscr{Y}$. The multiplicative function perspective on $\mathscr{Y}$ is therefore equivalent to the associated bundle perspective over X, providing an intrinsic groupoid-theoretic definition of the quantum states.

This concept of defining quantum states on the space of a groupoid resonates with ideas explored by Souriau later in his career concerning the role of positive type functions on the group of symmetries [Sou04] (which, in our case, is isomorphic to $\mathrm{Aut}(\mathbf{T}_\omega, \boldsymbol{\lambda})$).

The symmetry group $\mathrm{Diff}(\mathrm{X}, \omega) \simeq \mathrm{Aut}(\mathbf{T}_\omega, \boldsymbol{\lambda})$ acts naturally on $\mathscr{Y}$ via automorphisms $\Phi$. This action lifts to the space of multiplicative functions: $(\Phi_*\psi)(y) = \psi(\Phi^{-1}(y))$. Since $\Phi$ is a multiplicative map (a groupoid homomorphism), $\Phi_*\psi$ is also multiplicative, ensuring that the symmetry group acts on the candidate Hilbert space $\mathscr{H}$. As discussed in Article 17, this lift is an isomorphism, suggesting that central extensions may not arise at this level of representation, with the quantum nature encoded in the multiplicative property related to $\mathrm{T}_\omega$.

Developing the full geometric quantization program requires defining a suitable inner product on $\mathscr{H}$ and incorporating a notion of polarization to select physical states. These steps, along with a detailed study of the relevant representations of $\mathrm{T}_\omega$ and the induced representations of $\mathrm{Diff}(\mathrm{X}, \omega)$, constitute essential directions for future research stemming from this prequantization framework on diffeological spaces.

**26. Relation to Classical Central Extensions.**

It is natural to question how the isomorphism $\mathrm{Aut}(\mathbf{T}_\omega, \boldsymbol{\lambda}) \simeq \mathrm{Diff}(\mathrm{X}, \omega)$ relates to the well-known phenomenon of non-trivial central extensions arising in the classical geometric quantization of symmetries, even on simply connected symplectic manifolds. For instance, the prequantization of the standard symplectic structure on $\mathbf{R}^{2n}$ leads to the Heisenberg group extension for the translation group, and the prequantization of the sphere $\mathrm{S}^2$ leads to the spin group $\mathrm{Spin}(3) \simeq \mathrm{SU}(2)$ as a central extension of SO(3).

The key to understanding this lies in distinguishing the object to which the symmetries are lifted:

∗ In the **classical prequantum bundle picture**, central extensions emerge when considering the lift of a symmetry group (or its Lie algebra) acting on the base manifold X to the total space of the *prequantum principal* U(1)-*bundle* $P \to \mathrm{X}$, or equivalently, to the space of sections of the associated line bundle. The failure of the lifted Lie bracket to match the lift of the base Lie bracket results in a cocycle

valued in the Lie algebra of the fiber group ($\mathfrak{u}(1)$), leading to a central extension of the symmetry Lie algebra and group by $U(1)$.

$*$ In the **prequantum groupoid picture**, the isomorphism $\mathrm{Aut}(\mathbf{T}_\omega, \boldsymbol{\lambda}) \simeq \mathrm{Diff}(X, \omega)$ states that the full group of $\omega$-preserving diffeomorphisms $\phi$ lifts *isomorphically* to act as automorphisms $\Phi$ of the *entire prequantum groupoid* $(\mathbf{T}_\omega, \boldsymbol{\lambda})$ that preserve the prequantum 1-form $\boldsymbol{\lambda}$. The object being acted upon without central extension here is the groupoid's space of morphisms $\mathscr{Y} = \mathrm{Paths}(X)/\sim_\omega$ endowed with the form $\boldsymbol{\lambda}$.

The classical prequantum bundle $Y \to X$ (or, in our framework, the generalized prequantum bundle $\mathscr{Y}_x = \mathrm{Mor}(\mathbf{T}_\omega, x, *) \to X$, obtained by fixing a source point $x$) is *not* the same object as the full groupoid $\mathbf{T}_\omega$. The groupoid $\mathbf{T}_\omega$ encompasses all such principal bundles $\mathscr{Y}_x$ for varying $x$, and the arrows of $\mathbf{T}_\omega$ can be seen as isomorphisms between fibers of these bundles over different points.

The central extensions observed in the classical setting *will reappear* when one descends from the groupoid level to consider actions on structures derived from the groupoid over the base space X. For example, when considering the lifting of symmetries to the total space of the principal bundle $\mathscr{Y}_x \to X$ or, crucially, to the space of sections of an associated bundle over X (which would be the candidate quantum Hilbert space), the central extension structure related to the period group $T_\omega$ will indeed typically emerge.

Therefore, the result $\mathrm{Aut}(\mathbf{T}_\omega, \boldsymbol{\lambda}) \simeq \mathrm{Diff}(X, \omega)$ does not contradict classical findings. Instead, it provides a potentially deeper insight: the groupoid $(\mathbf{T}_\omega, \boldsymbol{\lambda})$ serves as a fundamental, extension-free prequantum object where the entire symmetry group $\mathrm{Diff}(X, \omega)$ is faithfully represented. The "quantic anomaly" manifested by central extensions in the classical picture is then understood as a phenomenon that arises when one projects or realizes this fundamental structure and its symmetries within the context of bundles or representation spaces built over the base space X. The essential quantic information, captured by the period group $T_\omega$, resides fundamentally in the isotropy group of the groupoid, waiting to express itself as an extension upon descent to structures defined over X. This perspective offers significant conceptual clarity, especially when dealing with singular spaces where classical bundle constructions are challenging.

**Note.** This phenomenon of symmetry breaking is reminiscent of passing from an equivariant two-point moment map $\psi(x, x')$ to a no longer equivariant one-point moment map $\mu(x) = \psi(x_0, x) + c$, which is the general solution of the equation $\psi(x, x') = \mu(x') - \mu(x)$. The one-point moment map is no longer equivariant under the linear coadjoint action but with the affine coadjoint action modified by the cocycle $\theta(g) = \psi(x_0, g(x_0)) + \Delta c$.

**27. Future Directions.**

This paper focuses on the construction of the prequantum groupoid for connected and simply connected diffeological spaces. This allowed us to elaborate the core principles and strategies without the technical complexities inherent in the non-simply connected case.

A primary direction for future research is to extend this construction to the case of *non-simply connected diffeological spaces*. In this general setting, the structure is expected to depend not only on the period group $P_\omega$ (which needs to be defined properly for general diffeological spaces) but also on the homotopy of X, captured by its fundamental group $\pi_1(X)$.

Our strategy for this generalization will directly leverage the results established in the present paper. The universal covering space $\pi : \tilde{X} \to X$, which always exists in diffeology [PIZ13, §5.15], is connected and simply connected. The closed 2-form $\omega$ lifts to a closed 2-form $\tilde{\omega} = \pi^*\omega$ on $\tilde{X}$. Since $(\tilde{X}, \tilde{\omega})$ is simply connected, the construction from this paper (Part I) yields a prequantum groupoid $\mathbf{T}_{\tilde{\omega}}$ with its prequantum 1-form $\tilde{\boldsymbol{\lambda}}$ on $\tilde{\mathscr{Y}} = \mathrm{Mor}(\mathbf{T}_{\tilde{\omega}})$. The fundamental group $\Gamma = \pi_1(X)$ acts as the group of deck transformations on $\tilde{X}$. Since deck transformations preserve $\tilde{\omega}$, $\Gamma$ acts as a subgroup of $\mathrm{Diff}(\tilde{X}, \tilde{\omega})$. As shown in Section IV, these symmetries lift to faithful automorphisms of the prequantum groupoid $(\mathbf{T}_{\tilde{\omega}}, \tilde{\boldsymbol{\lambda}})$. We anticipate that the prequantum groupoid $\mathbf{T}_\omega$ over X can be constructed as a suitable quotient of $\mathbf{T}_{\tilde{\omega}}$ by this action of $\pi_1(X)$. Exploring how this symmetry action of $\pi_1(X)$ on $(\mathbf{T}_{\tilde{\omega}}, \tilde{\boldsymbol{\lambda}})$ gives rise to distinct structures over the base space X is the central path for Part II. Building on our previous work on manifolds [PIZ95], which classified prequantum bundles by $\mathrm{Ext}(\pi_1(M)^{ab}, P_\omega)$, we anticipate that this investigation will clarify how the interplay between $\pi_1(X)$ and $P_\omega$ will lead to a classification of such derived prequantum structures over X.

The path-based nature of our construction suggests a deep connection to Feynman's path integral formulation of quantum mechanics. The relationship between geometric quantization and path integrals has been a subject of interest for many years among physicists, with various approaches explored in the literature (see, e.g., [Hor24] for recent reflections on prequantization from a path integral viewpoint). Further investigation into how the structure of the prequantum groupoid $\mathbf{T}_\omega$ directly encodes elements of the path integral formulation is a compelling direction for future research.

Beyond the diffeological generalization, several other avenues are opened by this work:

∗ Apply the prequantum groupoid construction to the coadjoint orbits of Lie groups, which are fundamental examples in geometric quantization. Investigate how their specific properties, such as transitivity and the existence of invariant metrics (for semi-simple groups), might simplify the structure of $(\mathbf{T}_\omega, \boldsymbol{\lambda})$ or provide new insights into the quantization of these spaces.

∗ Explore the space of multiplicative functions on the groupoid $\mathbf{T}_\omega$ as candidate wave functions. Investigate the representations of the symmetry group $\mathrm{Diff}(X, \omega)$ on this space, including the study of representations of discrete symmetry groups, which is naturally accommodated by the diffeological setting.

∗ How does the Maslov correction, crucial in the framework of geometric quantization, manifest in the space of paths and within the structure of the prequantum groupoid? In particular, in the case of the quantization of the harmonic oscillator (see, e.g., [Sou75]).

∗ How can Lagrangian polarization be incorporated into this path-space framework to include the reduction of the parasymplectic space to the space of configurations?

∗ Furthermore, the flexibility of the diffeological setting may provide a rigorous framework for understanding and justifying the use of *singular polarizations*, which have been considered by some authors but lack a fully developed mathematical foundation in the manifold setting.

These questions point towards the development of a full geometric quantization program within the diffeological setting, applicable to a wide range of spaces, including those with singularities.

## References

[DI83] Paul Donato and Patrick Iglesias. *Exemple de groupes différentiels : flots irrationnels sur le tore*. Preprint CPT-83/P.1524, Centre de Physique Théorique, Marseille, July 1983. Published in *Comptes Rendus de l'Académie des Sciences*, 301(4), Paris, 1985. `http://math.huji.ac.il/ piz/documents/EDGDFISLT.pdf`

[GIZ23] Serap Gürer and Patrick Iglesias-Zemmour. *Orbifolds as stratified diffeologies*. Differential Geometry and its Applications, 86:101969, 2023.

[GIZ25] Serap Gürer and Patrick Iglesias-Zemmour, *On Diffeology of Orbit Spaces*, (Preprint available at `http://math.huji.ac.il// piz/documents/ODOOS.pdf`).

[Hor24] Peter A. Horvathy. *Prequantisation from the path integral viewpoint*. Preprint arXiv:2402.17629, 2024.

[PIZ85] Patrick Iglesias. *Fibrés difféologiques et homotopie*. Thèse de doctorat d'état, Université de Provence, Marseille, 1985.

[PIZ95] Patrick Iglesias. *La trilogie du moment*. Ann. Inst. Fourier, 45(3):825–857, 1995.

[IKZ10] Patrick Iglesias, Yael Karshon, and Moshe Zadka. *Orbifolds as diffeologies*. Transactions of the American Mathematical Society, Vol. 362, Number 6, Pages 2811–2831, Providence R.I., 2010.

[PIZ10] Patrick Iglesias-Zemmour, *The moment maps in diffeology*, Mem. Amer. Math. Soc. **204** (2010), no. 962, viii+101 pp.

[PIZ13] Patrick Iglesias-Zemmour. *Diffeology*. Mathematical Surveys and Monographs, 185, Am. Math. Soc., Providence, RI, 2013. (Revised version by Beijing World Publishing Corp., Beijing, 2022.)

[PIZ22] Patrick Iglesias-Zemmour. *Diffeology*. Reprint by Beijing World Publishing Corporation, Ltd. Beijing Branch, 2022. ISBN 978-7-5192-9608-7. Accessible at `http://math.huji.ac.il/~piz/documents/Diffeology.pdf`

[PIZ15] ______ *Example of singular reduction in symplectic diffeology*. Proceedings of the American Mathematical Society, 144(3):1309–1324, 2016.

[IZP21] Patrick Iglesias-Zemmour and Elisa Prato. *Quasifolds, diffeology and noncommutative geometry*. J. Noncommut. Geom., 15(2):735–759, 2021.

[PIZ24] Patrick Iglesias-Zemmour, *Čech-de Rham bicomplex in diffeology*, Israel J. Math. **259** (2024), no. 1, 239–276.

[MW74] Jerrold E. Marsden and Alan Weinstein, *Reduction of symplectic manifolds with symmetry*, Rep. Mathematical Phys. **5** (1974), no. 1, 121–130.

[Sat56] Ichiro Satake, *On a generalization of the notion of manifold*, Proceedings of the National Academy of Sciences **42** (1956), 359–363.

[Sou70] Jean-Marie Souriau. *Structure des systèmes dynamiques*. Dunod, Paris, 1970.

[Sou75] ______ *Construction explicite de l'indice de Maslov. Applications*. In: Group theoretical methods in physics (Fourth Internat. Colloq., Nijmegen, 1975), Lecture Notes in Phys., Vol. 50, Springer, Berlin, 1976, pp. 117–148.

[Sou04] Jean-Marie Souriau, *C'est Quantique ? donc c'est géométrique*, in: Dominique Flament (dir), *Feuilletages -quantification géométrique : textes des journées d'étude des 16 et 17 octobre 2003, Paris, Fondation Maison des Sciences de l'Homme*, Série Documents de travail (Équipe F2DS), 2004.

EINSTEIN INSTITUTE OF MATHEMATICS, THE HEBREW UNIVERSITY OF JERUSALEM, CAMPUS GIVAT RAM, 9190401 ISRAEL.

*Email address*: `piz@math.huji.ac.il`

*URL*: `http://math.huji.ac.il/~piz`